\documentclass[twocolumn]{aastex63}
\usepackage{verbatim}
\usepackage{amsmath}
\let\tablenum\relax
\usepackage[detect-all]{siunitx}
\DeclareUnicodeCharacter{2212}{-}
\usepackage{xspace}
\usepackage{graphicx}
\newsavebox{\measurebox}
\usepackage[caption=false]{subfig}

\defcitealias{Harris1996}{H96} 	  	 
\newcommand{\harris}{\citetalias{Harris1996}\xspace}
\newcommand{\harrisp}{\citepalias{Harris1996}\xspace}

\received{}
\revised{}
\accepted{}

\submitjournal{ApJ}

\shortauthors{Prabhu et al.}

\graphicspath{{./}{}}

\begin{document}

\title{The First Extensive Exploration of UV-bright Stars in the Globular Cluster NGC 2808}

\correspondingauthor{Deepthi S. Prabhu}
\email{deepthi.prabhu@iiap.res.in}

\author[0000-0002-8217-5626]{Deepthi S. Prabhu}
\affiliation{Indian Institute of Astrophysics, Koramangala II Block, Bangalore-560034, India}
\affiliation{Pondicherry University, R.V. Nagar, Kalapet, 605014, Puducherry, India}
\author{Annapurni Subramaniam}
\affiliation{Indian Institute of Astrophysics, Koramangala II Block, Bangalore-560034, India}

\author[0000-0002-0801-8745]{Snehalata Sahu}
\affiliation{Indian Institute of Astrophysics, Koramangala II Block, Bangalore-560034, India}

\begin{abstract}
In this study, we identified and characterized the hot and luminous UV-bright stars in the globular cluster NGC 2808. We combined data from the Ultra Violet Imaging Telescope (UVIT) on-board the Indian space satellite, {\it AstroSat}, with the {\it Hubble Space Telescope} UV Globular Cluster Survey (HUGS) data for the central region (within $\sim$ $\ang[angle-symbol-over-decimal]{;2.7;} \times \ang[angle-symbol-over-decimal]{;2.7;}$) and {\it {\it Gaia}} and ground-based optical photometry for the outer parts of the cluster. We constructed the UV and UV-optical color-magnitude diagrams, compared the horizontal branch (HB) members with the theoretical zero-age HB and terminal-age HB models and identified 34 UV-bright stars. The spectral energy distributions of the UV-bright stars were fitted with theoretical models to estimate  their effective temperatures (12500 K - 100,000 K), radii (0.13 to 2.2 $R_{\odot}$), and luminosities ($ \sim 40$ to $3000$ $L_{\odot}$) for the first time. These stars were then placed on the H-R diagram, along with theoretical post-HB (pHB) evolutionary tracks to assess their evolutionary status. The models suggest that most of these stars are in the AGB-manqu\'e phase and all, except three, have evolutionary masses $<$ 0.53 ${\it M_{\odot}}$. We also calculated the theoretically expected number of hot post-(early)-AGB (p(e)AGB) stars in this cluster and found the range to match our observations. Seven UV-bright stars located in the outer region of the cluster, identified from the {\it AstroSat}/UVIT images, are ideal candidates for detailed follow-up spectroscopic studies. 
\end{abstract}


\section{Introduction}
Globular clusters (GCs) are ideal astrophysical laboratories to test the theories of stellar evolution, especially the late evolutionary stages of low mass stars. According to canonical stellar evolution models, the post-helium-core-burning (pHeCB) or post-horizontal branch (pHB) evolution of a star depends strongly on its envelope mass \citep{Dorman1993,Dorman1995}. After the depletion of helium (He) in the core, horizontal branch (HB) stars with the highest envelope masses enter the asymptotic giant branch (AGB) phase, undergo thermal pulsations, and end up losing their envelopes to become extremely hot stars with constant luminosity. Known as post-AGB (pAGB) stars, these are short-lived (lifetime $ < 10^{5}$ years) with luminosity, $ log {\it (L/L_{\odot})} \geq 3.1$. The HB stars with slightly lower envelope masses ($ > 0.02$ ${\it M_{\odot}}$) ascend the AGB but do not undergo thermal pulsations in this phase. They eventually lose their envelopes and evolve towards higher temperatures with slightly lower luminosities ($ log {\it (L/L_{\odot})}\sim$ 2.65 to 3.1) than the pAGB stars. These are known as post-early-AGB (peAGB) stars \citep{Brocato1990} and have a lifetime of $ \simeq 10^{5}$ years. After core-He exhaustion, HB stars with the lowest envelope masses cannot ascend the AGB and hence directly evolve towards the white dwarf cooling curve with a slight enhancement in luminosity and are referred to as AGB-manqu\'e stars \citep{GreggioRenzini1990}. The range of their luminosities is $log {\it (L/L_{\odot})} \sim $ 1.8 to 2.65,  with lifetimes  of 20 to 40 Myr \citep{Moehler2019}. These late phases of low-mass stellar evolution are the least understood owing to uncertainties in the C/O core size \citep{Charpinet2011, Constantino2015} and stellar winds in the red giant branch (RGB) phase which determine the envelope mass of HB stars \citep{Mcdonald2015, Salaris2016}. Due to the lack of 
detections of stars in these quickly evolving stages, the existing pHB models have not been tested extensively and have scope for improvement.  

A few pHeCB stars in various evolutionary stages are found in the ultraviolet (UV) images of GCs as highly prominent, luminous stars. Referred to as UV-bright stars \citep{Zinn1972}, these are observed to be brighter than the horizontal branch (HB) stars by one magnitude or more in the Far-UV (FUV) with FUV$-$Near UV (NUV) color $<$ 0.7 mag \citep{Schiavon2012}. 
Originally, these were defined as stars brighter than the HB and bluer than the RGB. They were looked for as stars whose {\it U}-band magnitudes were brighter than that of any other star in the GC (e.g., \cite{Zinn1972}). Since these searches were based on optical observations, they were biased towards the most luminous pAGB stars and were unable to detect the less luminous peAGB and AGB-manqu\'e stars \citep{Moehler2001}. Moreover, most of the stars detected were cooler than 30,000 K although there were theoretical predictions of even hotter pAGB stars. The crowded cores of GCs posed yet another difficulty for ground-based optical observations. 

The space missions, with capability to obtain images in the UV, opened a whole new arena for the study of hot stars in GCs. The cooler stellar populations such as main sequence (MS) and RGB stars are suppressed in the UV wavelengths, which helps to produce images with reduced stellar crowding in the central regions. UV-bright stars along with the hot extreme HB (EHB) stars contribute significantly to the UV luminosity of old stellar systems like GCs \citep{GreggioRenzini1990,GreggioRenzini1999,O'Connell1999}. Hence, these stars are speculated to be the reason for the UV-upturn phenomenon in elliptical galaxies \citep{GreggioRenzini1990,GreggioRenzini1999, Dorman1995,Brown1997,Brown2000}. UV study rather than optical is thus critical to perform an accurate evaluation of the contribution of hot stars in late evolutionary stages to the UV luminosity of old systems. For a detailed review of hot stars in GCs, see \citet{Moehler2001,Moehler2010}. \citet{Schiavon2012} presented a catalog of candidate pAGB, peAGB and AGB-manqu\'e stars in 44 Galactic GCs, including NGC 2808, using {\it Galaxy Evolution Explorer} ({\it GALEX}) FUV and NUV observations. However, this study was limited by the spatial resolution ($\sim$ $5\arcsec$) of GALEX. Also, the membership information for these stars were not available then.

\citet{Moehler2019} combined photometric observations from various missions such as Ultraviolet Imaging Telescope (UIT), {\it GALEX}, {\it Swift} Ultraviolet-Optical Telescope (UVOT) and {\it Hubble Space Telescope} ({\it HST}) to obtain the census of UV-bright stars in 78 GCs. The atmospheric parameters of the brightest pHB stars in the sample (including 3 stars from NGC 2808), derived from optical spectroscopic observations, were used to assess their evolutionary status. The number of theoretically predicted and observed hot pAGB stars for 17 GCs (excluding NGC 2808) were found to be comparable, though affected by poor statistics. The catalog of such stars in their list of clusters is yet to be published. Several optical spectroscopic studies of previously identified bright pHB stars have been performed to shed light on their chemistry and evolutionary status \citep{Thompson2007,Chayar2015,Dixon2017,Dixon2019}.

\citet{Jain2019} studied NGC 2808 using the data obtained from the  Ultraviolet Imaging Telescope (UVIT). They reported the detection of hot stars belonging to different classes such as EHB, blue HB (BHB), red HB (RHB), blue hook (BHk), pAGB, and blue straggler stars (BSSs), from UV color-magnitude diagrams (CMDs), albeit without membership analysis. They focused on the photometric gaps in the UV color-magnitude diagrams (CMDs) and the multiple stellar populations in the cluster.

In this work, we study the unexplored UV-bright member stars in NGC 2808. This is one of the massive and dense GCs with an age = $10.9 \pm 0.7$ Gyrs \citep{Massari2016}, [Fe/H] = $-$1.14 dex and located at a distance of 9.6 kpc \citep[2010 edition, H96]{Harris1996}. We combine the archival UV data from the UVIT with the UV-optical observations from the {\it HST} \citep{Brown2001, Nardiello2018,Piotto2015}, optical observations from {\it {\it Gaia}} \citep{Helmi2018} and ground-based telescopes \citep{Stetson2019}. The member stars in the inner (central region within $\sim$ $\ang[angle-symbol-over-decimal]{;2.7;} \times \ang[angle-symbol-over-decimal]{;2.7;}$, covered by {\it HST} WFC3/UVIS) and outer regions (the region outside {\it HST} field of view (FOV)) of the cluster are identified utilizing the proper motion-based membership information from the {\it HST} and {\it {\it Gaia}} catalogs, respectively. We present the catalog of UV-bright member stars in this cluster (from inner to outer regions) for the first time along with their surface parameters derived through the analysis of their spectral energy distributions (SEDs). We also investigate the evolutionary status of these UV-bright stars and compare the observed number of hot p(e)AGB \footnote{We use the notation p(e)AGB to indicate either pAGB or peAGB star.} stars with the expected number derived from theoretical estimations. Though \citet{Jain2019} used the UVIT data to study the HB population, here we aim to focus on the UV-bright members of the cluster and parameterize them to throw light on the short-lived late evolutionary phases of low mass stars.

The layout of the paper is as follows. Section \ref{obs} gives an account of the observations and the adopted data reduction procedure. The cross-match of the different datasets and the UV-optical CMDs are presented in Section \ref{CMDs}. Sections \ref{UV bright}, \ref{SEDs}, \ref{evolution tracks} describe the observed UV-bright stars, their SEDs and evolutionary phases, respectively. The results are discussed in Section \ref{discussion} and a summary is presented in Section \ref{summary}.


\section{Observations and Data Reduction} \label{obs}

This work utilizes data from the UVIT on-board {\it AstroSat}, India's first multiwavelength space observatory. The UVIT consists of twin telescopes, each of 38cm diameter, one dedicated for FUV ($\lambda = 1300-1800$ \AA), and the other for NUV ($\lambda = 2000-3000$ \AA) and VIS ($\lambda = 3200-5500$ \AA) 
pass bands. The VIS data is used for drift-correcting the images. The UVIT has a circular FOV of $28\arcmin$ diameter and consists of multiple filters in each pass band. The FUV and NUV detectors operate in photon-counting mode and the VIS detector in integration mode. Further details regarding the instrument and calibration can be found in \citet{Tandon2017}.

\begin{figure} [htbp]
\centering
    \includegraphics[width=0.45\textwidth]{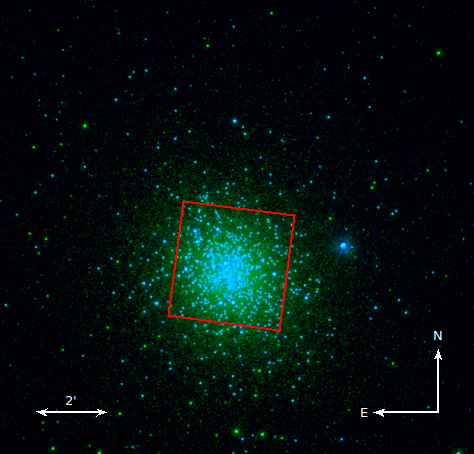}
    \caption{The UVIT image of NGC 2808 with F154W detections in blue and N242W in green. The {\it HST} WFC3/UVIS FOV which covers the inner $\sim$ $\ang[angle-symbol-over-decimal]{;2.7;} \times \ang[angle-symbol-over-decimal]{;2.7;}$ region of the cluster is marked in red.}
    \label{ngc2808_rgb}
\end{figure}

We used the archival UVIT data for NGC 2808, in two FUV (F154W and F169M) and four NUV filters (N242W, N245M, N263M and N279N). The images were created using the CCDLAB software package \citep{Postma2017} by correcting for the spacecraft drift, geometrical distortions and flat-fielding. The images created for each orbit were then aligned and merged to create the final science-ready image in each filter. The final exposure times for the science ready image in various filters are tabulated in Table~\ref{phot_details}. The UVIT image of NGC 2808 is shown in Figure~\ref{ngc2808_rgb} with blue corresponding to F154W detections and green to those in N242W. 

\subsection{Photometry}\label{photometry}

\begin{figure} [htb!]
\makebox[0.45\textwidth]{
    \includegraphics[width=0.58\textwidth]{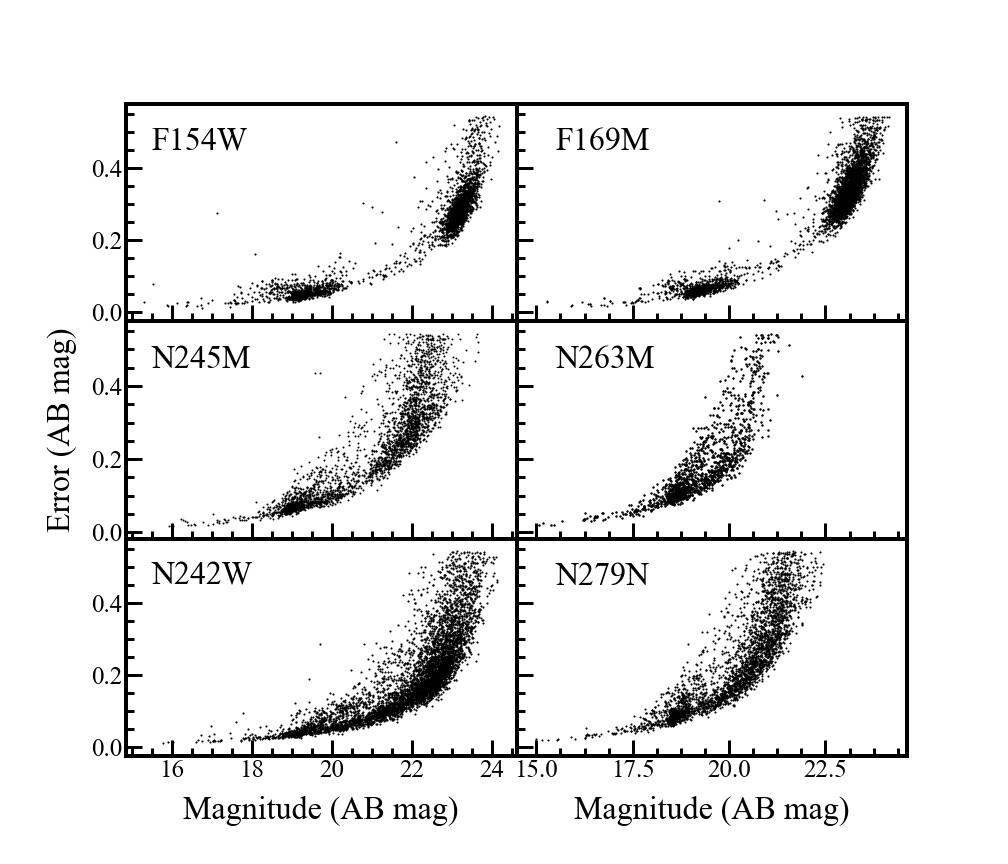}
    }
    \caption{The magnitude (without extinction correction) versus PSF-fit errors for the UVIT observations of NGC 2808 in six filters. The filter names are indicated in the top left corner of each panel.}
    \label{phot_err}
\end{figure}

Crowded-field photometry was performed on the UVIT images using DAOPHOT software package of IRAF/NOAO \citep{Stetson1987} through the following steps. Stars in the images were located using the DAOFIND task, and their aperture photometry magnitudes were computed using the PHOT task. The model point spread function (PSF) was generated using a few bright isolated stars in the image. It was then fitted to all the detected stars, and the PSF-fitted magnitudes were obtained using the ALLSTAR task. The curve-of-growth technique was used to calculate the aperture correction value, which was then applied to the magnitudes. To get the final magnitudes in each filter, saturation correction was applied according to the method described in \citet{Tandon2017}. The number of detected stars in various filters is presented in Table~\ref{phot_details}. The plots of magnitude versus PSF-fit error for different filters are shown in Figure~\ref{phot_err}. 

The magnitudes in all filters were corrected for extinction by adopting a reddening value, ${\it E(B-V)}$ = 0.22 mag \harrisp and the ratio of total to selective extinction, ${\it R_V}$ = 3.1. The corresponding extinction coefficient in the {\it V} band is, ${\it A_{V}}$ = 0.682 mag. The extinction coefficients presented in Table~\ref{phot_details} were calculated using the reddening law of \citet{Cardelli1989}.


\begin{table*}[htb!]
\caption{The UVIT observation and photometry details for NGC 2808}
\label{phot_details}
\makebox[0.86\linewidth]
{
\begin{tabular}{cccccccc} 
\toprule
Filter & $\lambda_{mean}$ & $\Delta\lambda$ & Zero point & Exp. time & No. of stars & FWHM of Model PSF & $A_{\lambda}$ \\
 & (\AA) & (\AA) & (mag) & (s) &  & ($arcsec$) & (mag) \\
 \hline
F154W & 1541 & 380 & 17.77 & 4987.34 & 2692 & 1.47 & 1.79 \\
F169M & 1608 & 290 & 17.45 & 4220.36 & 3996 & 1.45 & 1.75 \\
N242W & 2418 & 785 & 19.81 & 1040.99 & 5056 & 1.41 & 1.70 \\
N245M & 2447 & 280 & 18.50 & 886.42 & 2686 & 1.64 & 1.65 \\
N263M & 2632 & 275 & 18.18 & 354.46 & 1309 & 1.58 & 1.43 \\
N279N & 2792 & 90 & 16.50 & 2629.94 & 2868 & 1.42 & 1.32\\
\toprule
\end{tabular}
}
\end{table*}

\section{UV-optical and UV CMDs} \label{CMDs}

\subsection{Cross-match of the UVIT data with the {\it HST} data}

In order to identify various stars detected in the UVIT images, we cross-matched them with the {\it HST} UV Globular Cluster Survey (HUGS) catalog \citep{Nardiello2018,Piotto2015} in the inner region of the cluster (within $\sim$ $\ang[angle-symbol-over-decimal]{;2.7;} \times \ang[angle-symbol-over-decimal]{;2.7;}$). The HUGS astro-photometric catalog consists of data in the WFC3/UVIS F275W (NUV), F336W ({\it U}) and F438W ({\it B}) filters along with ACS/WFC F606W ({\it V}) and F814W ({\it I}) filters. The catalog also provides cluster membership probabilities estimated based on stellar proper-motions.  We selected stars with membership probability greater than $90\%$ as cluster members. As we are interested in the UV-bright population of this cluster, we chose stars which are expected to be bright in the UV, such as the pHB, HB and BSSs from the {\it HST} CMD and identified their unique counterparts in the UVIT images as explained below. 

\begin{figure*}[!htb]
\makebox[0.98\textwidth]
{
    \includegraphics[width=1.1\textwidth]{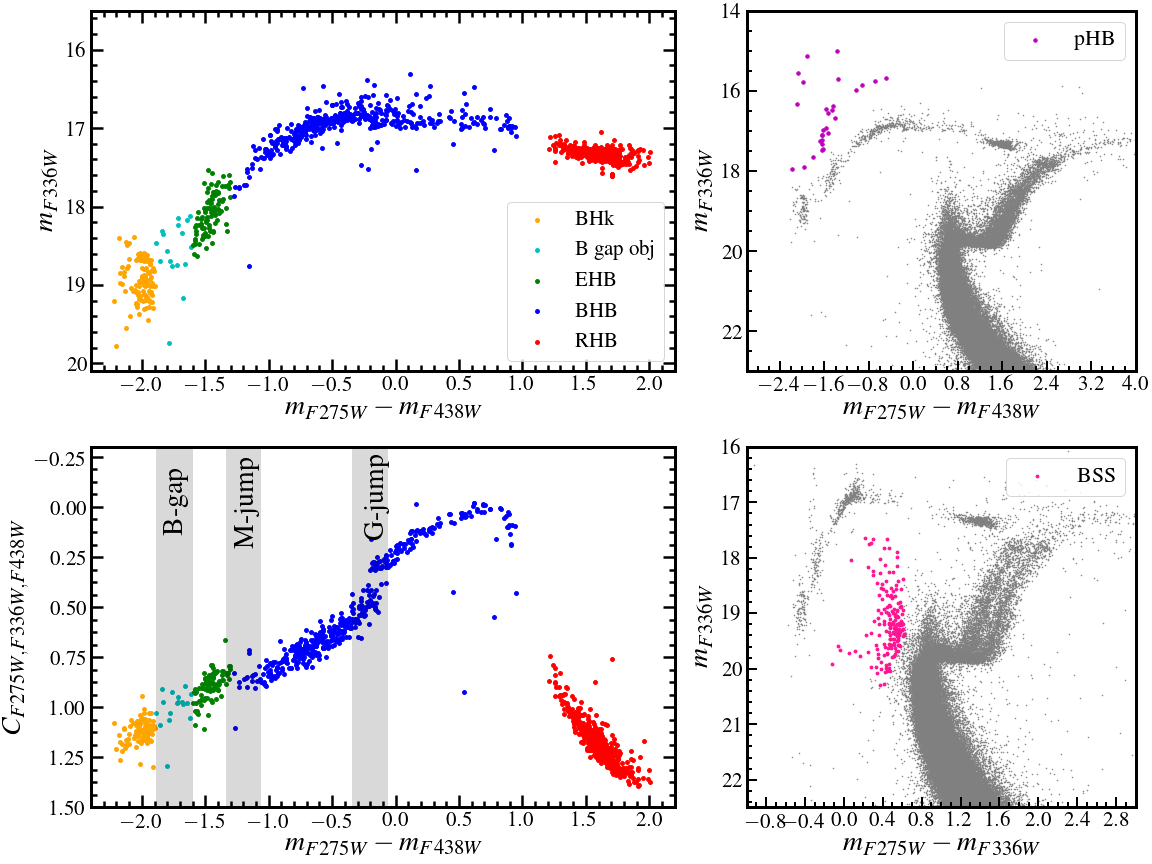}
}
    \caption{On the left panels are the CMD and CCP showing the categorization of HB stars into RHB, BHB, EHB, and BHk stars using HUGS data. The prominent photometric discontinuities such as the Grundahl jump (G-jump), the Momany jump (M-jump) and the gap between EHB and BHk stars (B-gap), are shown in the CCP. The top and bottom right panels show the CMDs used to select the pHB stars and BSSs respectively, with all the other member stars shown in gray.}
\label{selection_of_inner_stars}
\end{figure*}

\begin{figure}[ht]
\centering
    \includegraphics[width=0.3\textwidth]{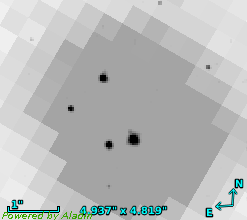}
    \caption{An example of a possible wrong cross-match between {\it HST} and UVIT detected stars due to the crowding in the inner region of the cluster. The UVIT detection in F154W filter is shown in gray, overlaid with the {\it HST} F275W detected stars in black.}
    \label{wrong_cm} 
\end{figure}

The CMDs and color-color plane (CCP) using the {\it HST} data for the inner region of the cluster are shown in Figure \ref{selection_of_inner_stars} where the stars are color-coded according to their classification. The top right panel shows the 28 identified pHB member stars. The bottom right panel shows the 176 identified BSS members as per the procedure described in \cite{Raso2017}. The division of HB into RHB, BHB, EHB, B gap objects and BHk was done using the ${\it m}_{F275W} - {\it m}_{F438W}$ vs. ${\it C}_{F275W,F336W,F438W}$ plane \citep{Brown2016}, as shown in the left panels. The classification adopted here is similar to that of \citet{Brown2016}. The RR Lyrae stars were identified by cross-matching with the data from \citet{Kunder2013}. Since the {\it HST} images have a much better spatial resolution, a simple cross-match with the UVIT data may yield wrong identifications due to crowding in the innermost region of the cluster. A case of multiple {\it HST} stars for a single UVIT detection is demonstrated in Figure \ref{wrong_cm}, where we have overlaid the {\it HST} F275W image and the UVIT F154W image. The gray patch is the UVIT detection, whereas in black are the stars are from {\it HST} in a region of about $5\arcsec \times 5\arcsec$. In order to avoid such wrong/multiple identifications, we used a python code which returns only those {\it HST} stars among the pHB, HB and BSS stars, which have no neighbors within $1.8\arcsec$ radius (maximum UVIT PSF) from them. As the crowding is maximum near the cluster center, no {\it HST}-UVIT cross-match was performed for the innermost region (radius $<$ $30\arcsec$). Finally, 491 stars were selected from the {\it HST} catalog, and these were then cross-matched with the UVIT detected stars with a maximum match radius of $1\arcsec$. All the cross-matched stars were found to have a photometric error $<$ 0.2 mag in all the UVIT filters.

\subsection{Cross-match of the UVIT data with the {\it {\it Gaia}} and ground-based optical data}

The UVIT FOV is much larger than the {\it HST} FOV, and it covers the outer region of the cluster as well. In order to effectively use the stars detected by UVIT in the outer region, the UVIT data were combined with other data for their identification and membership, as explained below. The list of possible member stars of the cluster in the outer region was obtained from the {\it {\it Gaia}} DR2 paper,  \citet{Helmi2018}, and their {\it UBVRI} photometry from \citet{Stetson2019}.
These datasets were cross-matched first to get the cluster members and their optical photometry. The resultant set was then cross-matched with the UVIT data, with a maximum match radius of $\ang[angle-symbol-over-decimal]{;;0.5}$. Then, only those stars with UVIT magnitude error $<$ 0.2  mag were included in the analysis. The cross-matched stars were visually checked for wrong/multiple identifications. Further, the $B$, $V$, and $I$ Johnson-Cousins magnitudes of all these stars were transformed into the equivalent {\it HST} filter magnitudes (WFC3/UVIS F438W, ACS/WFC F606W and ACS/WFC F814W), using the transformation equations of \citet{WEHarris2018} and \citet{Sirianni2005} so that stars in the inner and outer regions could be shown in the same color and magnitude plane. 


\subsection{Color-magnitude diagrams}
We constructed the UV-optical and UV CMDs for the inner and outer regions with the cross-matched members. The updated BaSTI (a Bag of Stellar Tracks and Isochrones) theoretical zero-age HB (ZAHB) and terminal-age HB (TAHB; end of He burning phase) models from \citet{Hidalgo2018} were fitted to the HB sequence of the cluster CMD. These models were generated in the UVIT, {\it HST}/WFC3 and {\it HST}/ACS filters by choosing metallicity [Fe/H] = $-$0.9 dex, He abundance = 0.249, solar scaled [$\alpha$/Fe] = 0.0,  and no convective overshoot. These models take into account the effects due to atomic diffusion at the hotter end of the HB as is evident from the close match with the CMDs. 

\begin{figure}[!htbp]
\centering
    \includegraphics[width=0.45\textwidth]{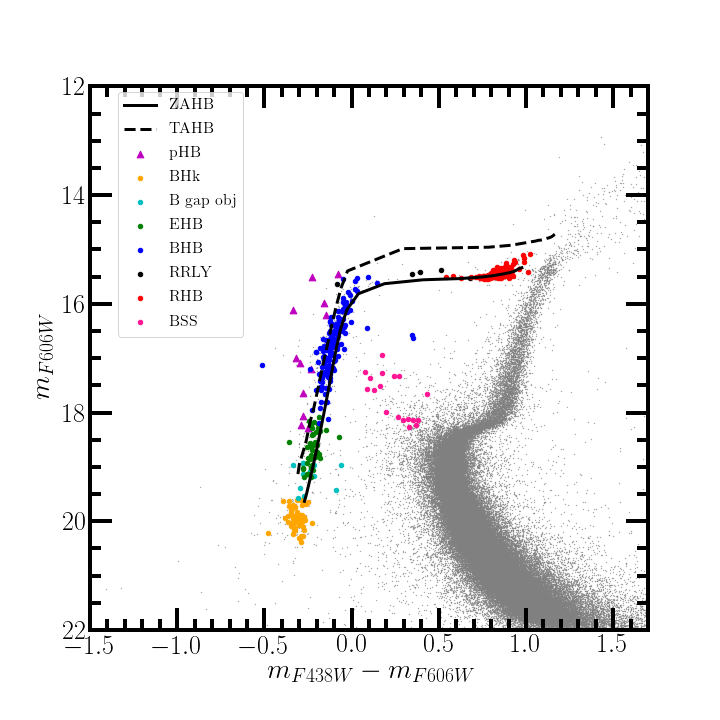}
    \caption{The optical CMD showing the 491 stars (in colored symbols) selected from {\it HST} catalog for cross-match with the UVIT data based on the selection criteria described in the text. All the cluster members (membership probability more than 90\%) in the {\it HST} FOV are shown with gray points and the black solid and dashed lines represents the BaSTI ZAHB and TAHB models with [Fe/H] = $-$0.9 dex, respectively.}
    \label{cross_match_inner}
\end{figure}

Figure \ref{cross_match_inner} shows the CMD of the cluster for the inner region using the {\it HST} data. Stars which are used for the cross-match with the UVIT data are shown along with the ZAHB and TAHB models. The gray dots represent stars with more than 90\% membership probability. 


\begin{figure*}[htbp]
\makebox[0.94\linewidth]
{
\includegraphics[width=1.05\textwidth]{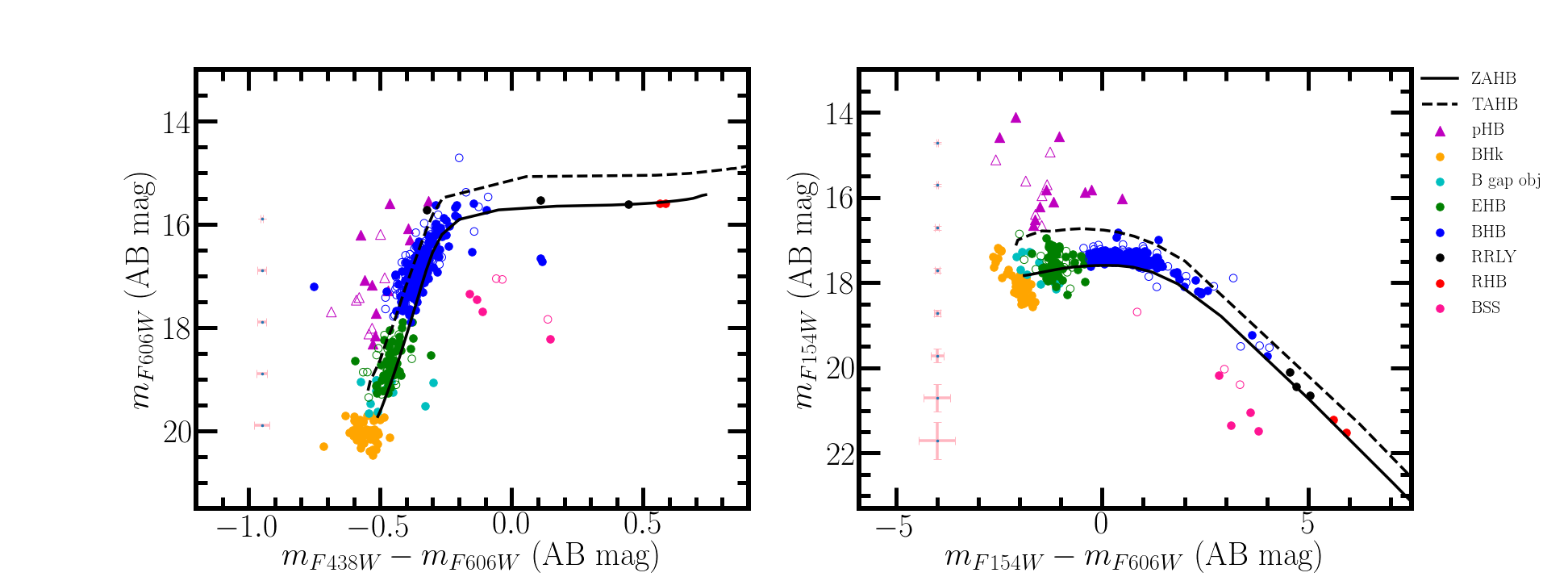}
}
\caption{The optical and FUV-optical CMDs for all the members in NGC 2808 common to the UVIT FUV F154W filter and other catalogs ({\it HST}, {\it Gaia} and ground-based optical data). The stars detected within the inner $\ang[angle-symbol-over-decimal]{;2.7;} \times \ang[angle-symbol-over-decimal]{;2.7;}$ region of the cluster are marked with filled symbols and those in the outer region with unfilled symbols. The black solid and dashed lines are the ZAHB and TAHB models same as in Figure \ref{cross_match_inner}. The photometric errors in magnitude and color are also shown along the left side of each plot.}
\label{FUV_cm_sources}
\end{figure*}

Figure \ref{FUV_cm_sources} shows the optical and FUV-optical CMDs for all the members detected in the F154W UVIT filter. To plot these CMDs, we converted the F438W and F606W magnitudes from the VEGA system to the AB system.\footnote{We used the conversion factors from \url{http://waps.cfa.harvard.edu/MIST/BC_tables/zeropoints.txt}.} The filled and unfilled symbols represent the stars located in the inner and outer regions of the cluster, respectively. In FUV, we have mainly detected the hotter part of the HB (BHB and EHB), BHk, pHB and a few BSSs. Three RR Lyrae variables are also identified. In the ${\it m}_{F154W} - {\it m}_{F606W}$ vs. ${\it m}_{F154W}$ plot (shown in the right panel), the BHB stars have a relatively tight distribution in magnitude, whereas the EHB, B gap and BHk stars show a spread in the ${\it m}_{F154W}$ magnitude, and with respect to the ZAHB track. Many of them are also found to be fainter than the ZAHB model. One of the detected BSSs, located in the outer region, is very bright in F154W filter. The RHB stars are hardly detected as they are not hot enough to emit in FUV wavelengths. The detected pHB stars are found to be brighter than the TAHB with a spread of about 2 magnitudes in F154W filter and 4 magnitudes in the FUV-optical color. We discuss these UV-bright stars in detail in the next section.

\begin{figure*}[htb!]
\makebox[0.94\linewidth]
{
\includegraphics[width=1.05\textwidth]{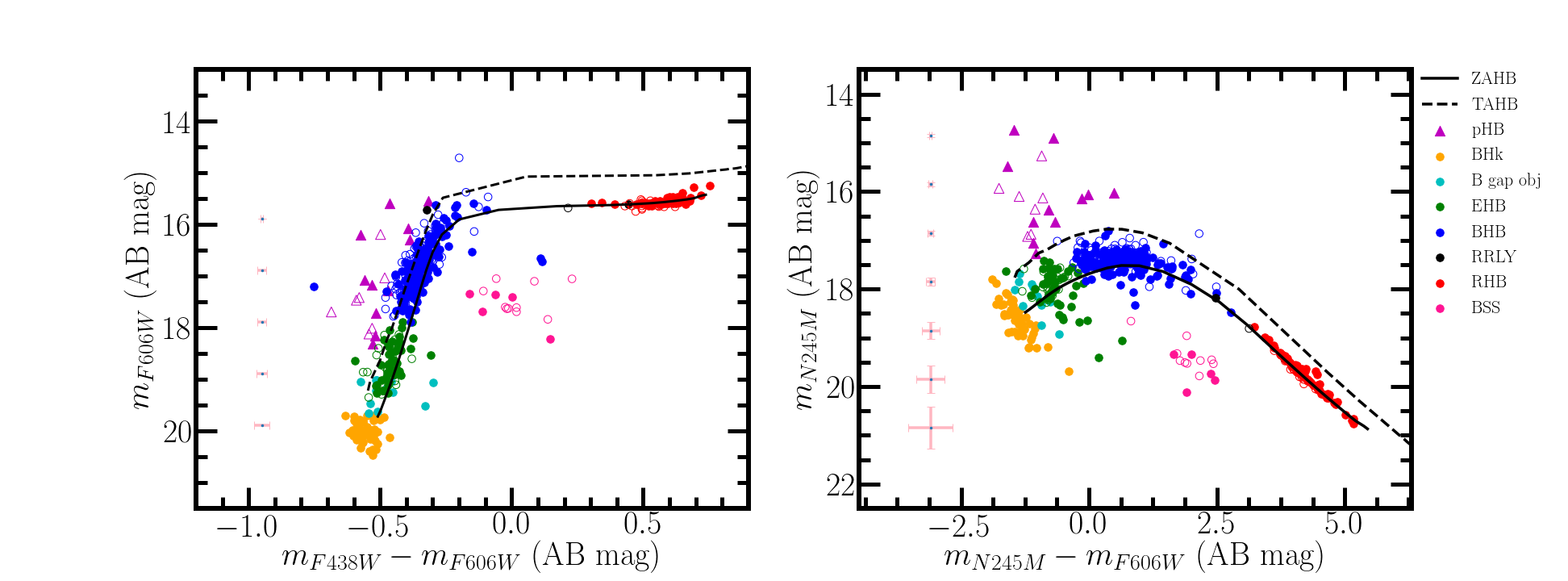}
}
\caption{The optical and NUV-optical CMDs for all the stars in NGC 2808 common to the UVIT NUV N245M filter and other catalogs ({\it HST}, {\it Gaia} and ground-based optical data). The details are same as in Figure \ref{FUV_cm_sources}. }
\label{NUV_cm_sources}
\end{figure*}

Similarly, Figure \ref{NUV_cm_sources} shows the optical and NUV-optical CMDs for stars detected in the N245M UVIT filter. In the N245M filter, apart from the above-mentioned sequences, we have detected the cooler RHB population as well. The RHB stars form a tight sequence when compared to the rest of the HB sequence, and are located close to the ZAHB track. On the other hand,  the other HB phases show a large spread, with many of them located above and below the ZAHB track. The BHk sequence has the largest statistically significant spread in the N245M magnitude, as is evident from the ${\it m}_{N245M} - {\it m}_{F606W}$ vs. ${\it m}_{N245M}$ CMD (right panel). The bright BSS is found to be located slightly fainter than the faintest BHB star. 

\begin{figure*}[htb]
\makebox[0.94\linewidth]
{
\includegraphics[width=1.05\textwidth]{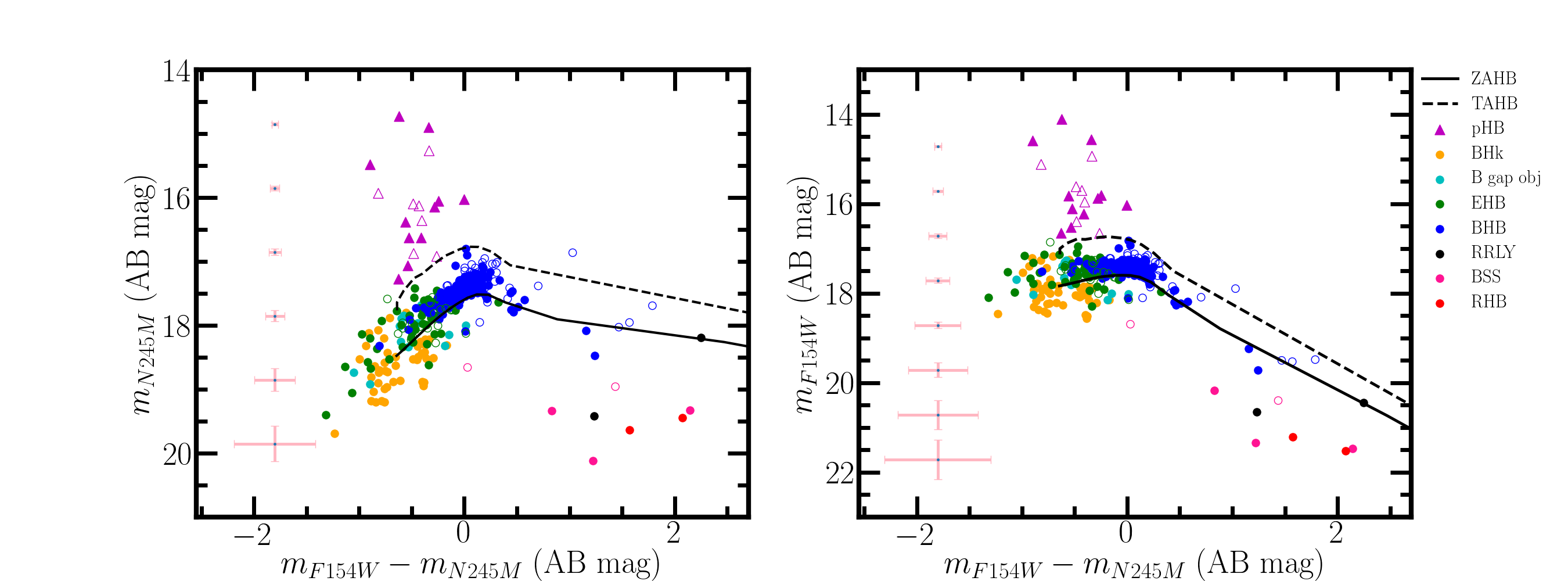}
}
\caption{The UV CMDs for the cluster member stars common in F154W and N245M filters. The UV-bright stars (shown as purple triangles) are clearly brighter than the ZAHB by 1 mag or more in FUV and have FUV$-$NUV $<$ 0.7 mag.}
\label{baf2_b13_cmds}
\end{figure*}

The UV CMDs for all the stars common in the UVIT filters F154W and N245M are shown in Figure \ref{baf2_b13_cmds}. In the ${\it m}_{F154W}-{\it m}_{N245M}$ vs ${\it m}_{N245M}$ CMD (left panel), the HB stars show a progressive reduction in the N245M magnitude as a function of ${\it m}_{F154W}-{\it m}_{N245M}$ color. In the case of the ${\it m}_{F154W}-{\it m}_{N245M}$ vs ${\it m}_{F154W}$ CMD, the HB stars with color $<$ 0.0 mag, have a horizontal sequence. The hot BHB, EHB, B gap and BHk stars get mixed up in these CMDs. 
The bright BSS is found slightly fainter than the BHB stars, with ${\it m}_{F154W}-{\it m}_{N245M}$ color $\sim$ 0.0 mag. 

Table \ref{tab:no of stars} tabulates the number of stars belonging to different categories, detected in each UVIT filter. In the outer region, since cluster members were selected based on {\it Gaia} data with an approximate limiting magnitude of $V$ $\simeq$ 20.5 mag, we have zero stars belonging to the optically faint B gap and BHk categories. We plan to perform a separate detailed study of the HB stars and BSSs presented in this section, in the near future.


\begin{table*}[htbp]
\caption{Number of pHB stars, BSSs and different categories of HB stars in NGC 2808 detected in each UVIT filter in the region covered by {\it HST}. In parentheses are the numbers of these stars detected in the outer region of the cluster.}
\label{tab:no of stars}
\makebox[0.86\linewidth]
{
\begin{tabular}{ccccccccc} 
\toprule
Filter	&	$N_{pHB}$	&	$N_{BHk}$	&	$N_{Bgap}$	& $N_{EHB}$	& $N_{BHB}$ & $N_{RR Lyrae}$  & $N_{RHB}$ & $N_{BSS}$	\\
\hline
F154W & 11(7) & 53(0) & 12(0) & 48(21) & 147(65) & 3(0) & 2(0) & 4(3) \\
F169M & 11(7) & 52(0) & 12(0) & 49(21) & 146(64) & 2(1) & 1(0) & 3(1)\\
N242W & 11(7) & 49(0) & 11(0) & 45(21) & 146(66) & 4(1) & 119(124) & 9(28)\\
N245M & 11(7) & 48(0) & 12(0) & 45(20) & 146(66) & 2(1) & 65(23) & 5(10)\\
N263M & 11(7) & 23(0) & 8(0) & 29(17) & 140(66) & 3(1) & 38(4) & 4(1)\\
N279N & 11(7) & 40(0) & 10(0) & 40(17) & 147(64) & 5(1) & 103(49) & 8(2)\\
\toprule
\end{tabular}
}
\end{table*}

\section{OBSERVED UV-bright stars} \label{UV bright}
The main aim of this study was to identify and characterize the pHB stars in the inner and outer regions of the cluster. 
Among the UVIT detected stars in the outer regions, some bright stars 
were found to be non-members as per the list of possible members from \citet{Helmi2018}. We re-assessed the membership of these stars in the outer region, using methods presented in \citet{Singh2020} which also uses the {\it Gaia} DR2 data. Two stars were found to have a membership probability of $\sim$ 90\%. One star was found to have $\sim$ 40\% membership probability and this was studied by \citet{Moehler2019} using optical spectroscopy. We included these 3 stars along with the other pHB stars, for further study.

\begin{figure*} [!htb]
\centering
    \includegraphics[width=0.2\textwidth]{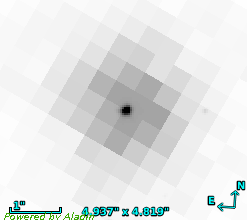}
    \includegraphics[width=0.2\textwidth]{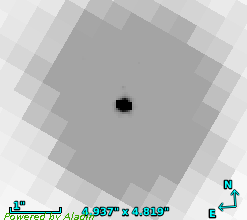}
    \includegraphics[width=0.2\textwidth]{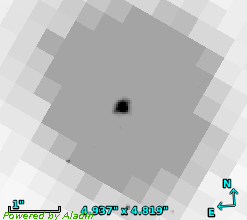}
    \includegraphics[width=0.2\textwidth]{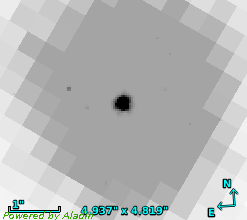}
    \\
    \includegraphics[width=0.2\textwidth]{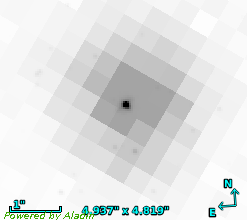}
    \includegraphics[width=0.2\textwidth]{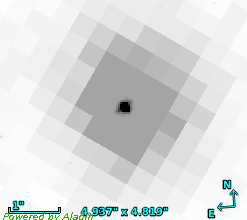}
    \includegraphics[width=0.2\textwidth]{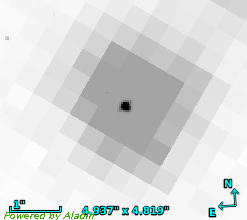}
    \includegraphics[width=0.2\textwidth]{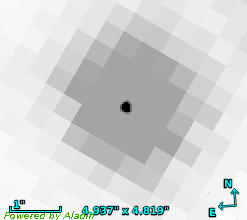}
    \\[\smallskipamount]
    \includegraphics[width=0.2\textwidth]{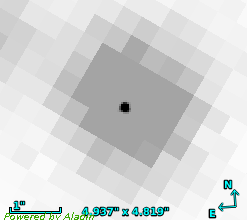}
    \includegraphics[width=0.2\textwidth]{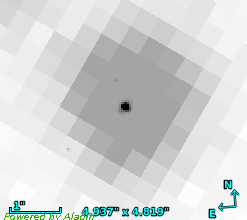}
    \includegraphics[width=0.2\textwidth]{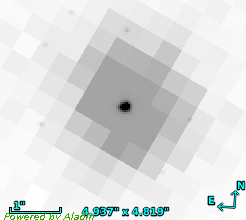}
        \\[\smallskipamount]
       \caption{The UVIT F154W images of all the 11 UV-bright stars in the inner $\ang[angle-symbol-over-decimal]{;2.7;} \times \ang[angle-symbol-over-decimal]{;2.7;}$ region of the cluster which are uniquely cross-matched with stars in the HUGS catalog. Here, the UVIT F154W image of each star is shown in gray and is overlaid with the corresponding {\it HST} F275W detection in black.}
\label{fig:UV-bright stars}
\end{figure*}

From our analysis, we find a total of 34 UV-bright member stars in this cluster, which could be classified as pHB stars. Of these, 18 stars have FUV, NUV flux measurements from the UVIT and are marked with purple triangles in the UV-optical and UV CMDs in Figures \ref{FUV_cm_sources}, \ref{NUV_cm_sources} and \ref{baf2_b13_cmds}. All these stars satisfy the observational criteria for them to be classified as UV-bright stars, as laid out in \citet{Schiavon2012}, i.e., they are brighter than the ZAHB by more than 1 mag in FUV and the FUV$-$NUV color is less than 0.7 mag. The UVIT magnitudes and magnitude errors for these stars are presented in the appendix (Table \ref{pHB UVIT mag}). Figure \ref{pHB_spatial_location} shows the spatial locations of all the 34 observed UV-bright stars marked over the UVIT F154W image. Here, 11 out of the 18 UVIT detected pHB stars (shown in red circles) lie in the {\it HST} FOV and they have been uniquely cross-matched. This can be seen in Figure \ref{fig:UV-bright stars}, where we have overlaid the {\it HST} F275W and the UVIT F154W images for these stars. We also note that similar unique cross-matches are achieved for all the stars shown in the CMDs. The remaining 16 UV-bright stars (marked in blue circles in Figure \ref{pHB_spatial_location}) are located in the crowded innermost region of the cluster and hence, could not be resolved by the UVIT. Among these, only 4 stars have FUV, NUV photometric measurements from the {\it HST} Space Telescope Imaging Spectrograph (STIS) instrument \citep{Brown2001}, whereas all the 16 have the NUV-optical data from HUGS catalog. 

These multi-wavelength photometric data for the UV-bright stars were used to determine their evolutionary status by estimating various parameters.

\begin{figure} [!htb]
\makebox[0.96\linewidth]
{
     \subfloat{%
       \includegraphics[width=0.53\textwidth]{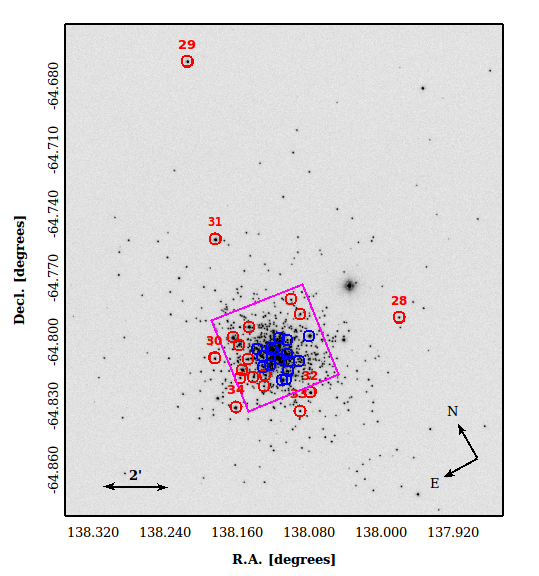}
     }
     }
     \caption{The 34 UV-bright stars in NGC 2808 marked over the UVIT F154W filter image of the GC. The magenta region marks the {\it HST} FOV. The star IDs are indicated near each star in the region outside the {\it HST} FOV. The 18 UV-bright stars with UVIT photometry are marked with red circles and the rest with blue circles.} 
     \label{pHB_spatial_location}
   \end{figure}
   

\section{Spectral energy distributions} \label{SEDs}

In order to estimate the parameters such as effective temperature (${\it T_{eff}}$), luminosity ($L$), and radius ($R$) of the UV-bright stars, their SEDs were constructed with the available photometric data points. For this purpose, a virtual observatory functionality, VOSA (VO SED Analyser; \citealt{Bayo2008}) was used. VOSA generates synthetic photometry for the chosen theoretical models using the transmission curves of the required photometric filters. The best-fit parameters of the SEDs were estimated by comparing the observed and synthetic photometric points using a $\chi^{2}$ minimization method. The $\chi_{red}^{2}$ value is calculated using the relation,

\begin{center}
\begin{equation}
\chi_{red}^{2} = \displaystyle \frac{1}{N-N_{f}} \sum_{i=1}^{N} \Bigg\{ \frac{(F_{o,i}-M_{d}F_{m,i})^{2}}{\sigma_{o,i}^{2}}\Bigg\}
\end{equation}
\end{center}

where {\it N} is the number of photometric points, ${\it N_{f}}$ is the number of fitted parameters for the model, ${\it F_{o,i}}$ is the observed flux, ${\it F_{m,i}}$ is the theoretical flux predicted by the model, ${\it M_{d}} = {\it (\frac{R}{D})^{2}}$ is the multiplicative dilution factor (where $R$ is the radius of the star and {\it D} is the distance to the star) and ${\it \sigma_{o,i}}$ is the error in the observed flux. We assumed a distance of {\it D} = 9.6 kpc and $E(B-V)$ = 0.22 mag \harrisp , for all the stars in the cluster. 
VOSA uses the Fitzpatrick reddening relation \citep{Fitzpatrick1999} to account for the extinction in the observed photometric points. 

For all the UV-bright stars except two, we used the Kurucz stellar atmospheric models \citep{Castelli1997, Castelli2003} and  log $\textit{g}$, [Fe/H] and ${\it T_{eff}}$ are the possible free parameters for fitting the SED. In these models, the range of admissible values for the free parameters are 0.0 to 5.0 for log $\textit{g}$, $-$2.5 to 0.5 for [Fe/H] and 3500 to 50000 K for $ {\it T_{eff}}$. For the remaining two stars, i.e., Star 3  and Star 32 in Table \ref{tab:sed params}, the T\"ubingen NLTE Model Atmosphere Package (TMAP T\"ubingen; \cite{Werner1999,Werner2003,Rauch2003}) was used as the Kurucz model parameter space was inadequate to fit the SEDs of these stars. The free parameters in these models include ${\it T_{eff}}$ with range 30000 to 1000000 K, log $\textit{g}$ ranging from 3.8 to 9, and H, He mass fractions in the range 0 to 1. 

To fit SEDs using Kurucz models, we fixed the [Fe/H] value at $-$1.0 dex, close to the value for the cluster. Additionally, we constrained the log $\textit{g}$ to range from 3 to 5 for Kurucz models and 3.8 to 6 for the TMAP models. This is the observed range of log $\textit{g}$ values for stars in pHB evolutionary phases from previous studies \citep{Moehler2019}. The important parameters obtained from SED analysis are tabulated in Table \ref{tab:sed params}. For the two stars fitted with TMAP models, the best-fit values of H, He mass fractions, respectively, are : 0.7383, 0.2495 for Star 3 and 0.8, 0.2 for Star 32. However, we note that SED fitting is not the optimal method to estimate the above parameters. We do not quote the log $\textit{g}$ values obtained from the SED fits as the SED fitting technique does not provide accurate values of this parameter.


\begin{table*}[htb!]
\caption{Table showing the results from the SED analysis of UV-bright stars in the cluster. Columns 1, 2 and 3 show the ID, R.A. and Decl. of the objects, respectively. The model used for the SED fit is shown in Column 4. The estimated parameters such as temperature, luminosity and radius (in solar units) of the stars and the errors in these parameters are tabulated in Columns 5 to 10. The errors in the parameters are derived as half the grid step, around the best-fit value. Columns 11 to 13, respectively, show the reduced chi square value corresponding to the fit, the number of photometric points  used for fitting and the corresponding photometric bands. Stars 1 to 11 have photometry from the UVIT and HUGS. Stars 12 to 15 have data points from the {\it HST} STIS and HUGS. Stars 16 to 27 have photometry only from HUGS. Stars 28 to 34 have photometric points from UVIT, {\it GALEX}, {\it Gaia} and ground-based optical data. In the last column, opt. stands for optical.}
\makebox[0.79\linewidth]
{
\begin{tabular}{ccccccccccccc}
\toprule
ID & R.A. & Decl. & Model & ${\it T_{eff}}$ & ${\it \Delta T_{eff}}$ & ${\it \frac{L}{L_{\odot}}}$ & ${\it \Delta \frac{L}{L_{\odot}}}$ & ${\it \frac{R}{R_{\odot}}}$ & ${\it \Delta \frac{R}{R_{\odot}}}$ & $\chi_{red}^{2}$ & ${\it N_{fit}}$ & Phot. used \\\
 & (deg) & (deg) &  & (K) & (K) &  &  &  &  &  &  & \\
 \toprule
Star 1  & 137.96181 & -64.84165 & Kurucz & 27000 & 500  & 50.80   & 0.27 & 0.33 & 0.01 & 7.58 & 11 & FUV, NUV, opt. \\
Star 2  & 138.02575 & -64.84307 & Kurucz & 21000 & 500  & 352.95  & 1.53 & 1.42 & 0.07 & 8.21 & 10 & FUV, NUV, opt. \\
Star 3  & 138.06046 & -64.86146 & TMAP   & 80000 & 3750 & 2857.45 & 0.88 & 0.28 & 0.03 & 3.31 & 8  & FUV, NUV, opt. \\
Star 4  & 138.05040 & -64.84379 & Kurucz & 39000 & 500  & 781.10  & 0.95 & 0.61 & 0.02 & 4.51 & 9  & FUV, NUV, opt. \\
Star 5  & 138.04719 & -64.87492 & Kurucz & 27000 & 500  & 59.54   & 0.25 & 0.35 & 0.01 & 2.34 & 10 & FUV, NUV, opt. \\
Star 6  & 138.06813 & -64.86459 & Kurucz & 12500 & 125  & 108.50  & 1.11 & 2.20 & 0.04 & 6.86 & 11 & FUV, NUV, opt. \\
Star 7  & 138.05264 & -64.86779 & Kurucz & 22000 & 500  & 85.73   & 0.44 & 0.64 & 0.03 & 3.36 & 9  & FUV, NUV, opt. \\
Star 8  & 138.04842 & -64.84887 & Kurucz & 16000 & 500  & 117.20  & 0.96 & 1.41 & 0.09 & 6.44 & 10 & FUV, NUV, opt. \\
Star 9  & 138.04760 & -64.85790 & Kurucz & 17000 & 500  & 109.80  & 0.78 & 1.21 & 0.07 & 3.50 & 10 & FUV, NUV, opt. \\
Star 10 & 138.03820 & -64.87028 & Kurucz & 22000 & 500  & 106.20  & 0.62 & 0.71 & 0.03 & 5.78 & 10 & FUV, NUV, opt. \\
Star 11 & 137.96088 & -64.85092 & Kurucz & 26000 & 500  & 79.93   & 0.32 & 0.44 & 0.02 & 3.94 & 9  & FUV, NUV, opt. \\
Star 12 & 138.02716 & -64.86694 & Kurucz & 25000 & 500  & 102.60  & 0.87 & 0.54 & 0.02 & 0.10 & 7  & FUV, NUV, opt. \\
Star 13 & 138.01470 & -64.85891 & Kurucz & 15000 & 500  & 128.83  & 2.02 & 1.68 & 0.11 & 0.09 & 7  & FUV, NUV, opt. \\
Star 14 & 138.01349 & -64.86046 & Kurucz & 27000 & 500  & 86.65   & 0.09 & 0.42 & 0.02 & 4.86 & 7  & FUV, NUV, opt. \\
Star 15 & 138.01173 & -64.86674 & Kurucz & 26000 & 500  & 131.15  & 0.22 & 0.56 & 0.02 & 1.59 & 7  & FUV, NUV, opt. \\
Star 16 & 138.03578 & -64.86580 & Kurucz & 23000 & 500  & 54.90   & 0.04 & 0.47 & 0.02 & 1.39 & 5  & NUV, opt.      \\
Star 17 & 138.03122 & -64.85554 & Kurucz & 25000 & 500  & 41.40   & 0.02 & 0.34 & 0.01 & 8.76 & 5  & NUV, opt.      \\
Star 18 & 138.03010 & -64.86050 & Kurucz & 22000 & 500  & 211.00  & 0.07 & 1.00 & 0.05 & 6.01 & 5  & NUV, opt.      \\
Star 19 & 138.02244 & -64.87728 & Kurucz & 45000 & 500  & 633.00  & 0.10 & 0.41 & 0.01 & 1.33 & 5  & NUV, opt.      \\
Star 20 & 138.01776 & -64.87794 & Kurucz & 27000 & 500  & 72.00   & 0.03 & 0.39 & 0.01 & 1.98 & 5  & NUV, opt.      \\
Star 21 & 138.01103 & -64.87463 & Kurucz & 24000 & 500  & 49.30   & 0.08 & 0.41 & 0.02 & 3.75 & 5  & NUV, opt.      \\
Star 22 & 138.00244 & -64.87014 & Kurucz & 22000 & 500  & 102.00  & 0.69 & 0.69 & 0.03 & 0.02 & 5  & NUV, opt.      \\
Star 23 & 138.00133 & -64.86623 & Kurucz & 50000 & 500  & 113.00  & 0.02 & 0.14 & 0.00 & 4.61 & 5  & NUV, opt.      \\
Star 24 & 137.99932 & -64.85673 & Kurucz & 50000 & 500  & 97.80   & 0.13 & 0.13 & 0.00 & 3.77 & 5  & NUV, opt.      \\
Star 25 & 137.99187 & -64.87307 & Kurucz & 35000 & 500  & 56.10   & 0.18 & 0.20 & 0.01 & 0.33 & 5  & NUV, opt.      \\
Star 26 & 137.99182 & -64.85956 & Kurucz & 27000 & 500  & 67.70   & 0.32 & 0.38 & 0.01 & 0.11 & 5  & NUV, opt.      \\
Star 27 & 137.96443 & -64.86377 & Kurucz & 50000 & 500  & 988.00  & 1.22 & 0.42 & 0.01 & 0.26 & 5  & NUV, opt.      \\
Star 28 & 137.85352 & -64.87929 & Kurucz & 26000 & 500  & 65.36   & 0.97 & 0.39 & 0.02 & 2.58 & 15 & FUV, NUV, opt. \\
Star 29 & 137.92453 & -64.70158 & Kurucz & 32000 & 500  & 145.80  & 1.09 & 0.39 & 0.01 & 2.66 & 15 & FUV, NUV, opt. \\
Star 30 & 138.08369 & -64.84865 & Kurucz & 25000 & 500  & 98.59   & 1.03 & 0.52 & 0.02 & 2.89 & 13 & FUV, NUV, opt. \\
Star 31 & 138.00728 & -64.79293 & Kurucz & 23000 & 500  & 251.00  & 2.12 & 0.99 & 0.04 & 4.49 & 14 & FUV, NUV, opt. \\
Star 32 &  137.99974 & -64.89071 & TMAP	& 100000 & 5000 & 3010.25	& 1.20 & 0.18 & 0.02 & 9.06 & 12 & FUV, NUV, opt. \\
Star 33 & 138.02306 & -64.89646 & Kurucz & 26000 &	500	& 53.26	& 0.51 & 0.36 & 0.01 & 2.39 & 11 &	FUV, NUV, opt. \\
Star 34 & 138.09187 & -64.87727 & Kurucz &	24000 & 500 & 119.91 & 0.75 & 0.63 & 0.03 &	4.40 & 11 & FUV, NUV, opt. \\
\toprule
\end{tabular}
}
\label{tab:sed params}
\end{table*}


The SEDs for stars 1 to 11 contain the UVIT (FUV, NUV) and the {\it HST} (NUV, optical) photometric data points. Stars 12 to 26, lying within the inner region of the cluster, could not be resolved by the UVIT. Hence, only the {\it HST} photometry was used to construct the SEDs of these stars. Four stars among these, namely, Stars 12, 13, 14 and 15, have the {\it HST} STIS photometry (FUV, NUV) from \citet{Brown2001} apart from the NUV and optical HUGS data. For stars 27 to 34, which lie outside the {\it HST} FOV, we utilize the UVIT, {\it GALEX} \citep{Schiavon2012}, {\it Gaia} and ground-based optical data for SED generation. Examples of SED fits for star 3 and star 27 are shown in Figure \ref{SED_eg}. The SEDs for the other UVIT-resolved stars are shown in the appendix (Figure \ref{seds_uvit_phb}). The residuals shown in these plots were calculated for each data point as follows :

\begin{center}
\begin{equation}
Residual = \frac{F_{o} - F_{m}}{F_{o}}
\end{equation}
\end{center}

where $F_{o}$ and $F_{m}$ are the observed and model fluxes corresponding to the photometric points.


\begin{figure*}[!htb]
\makebox[\linewidth]
{
     \subfloat{%
         \includegraphics[width=0.55\textwidth]{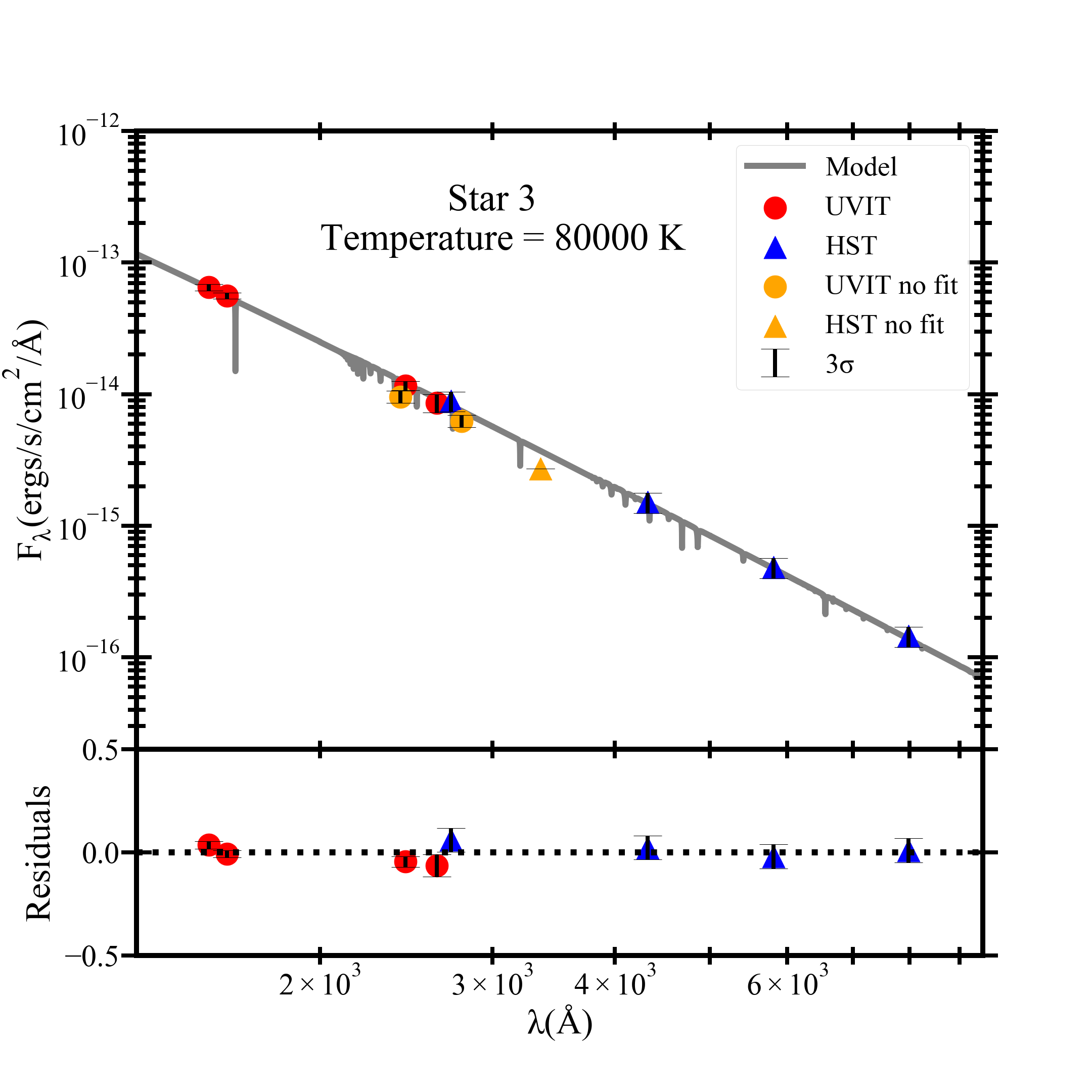}}
    \subfloat{%
         \includegraphics[width=0.55\textwidth]{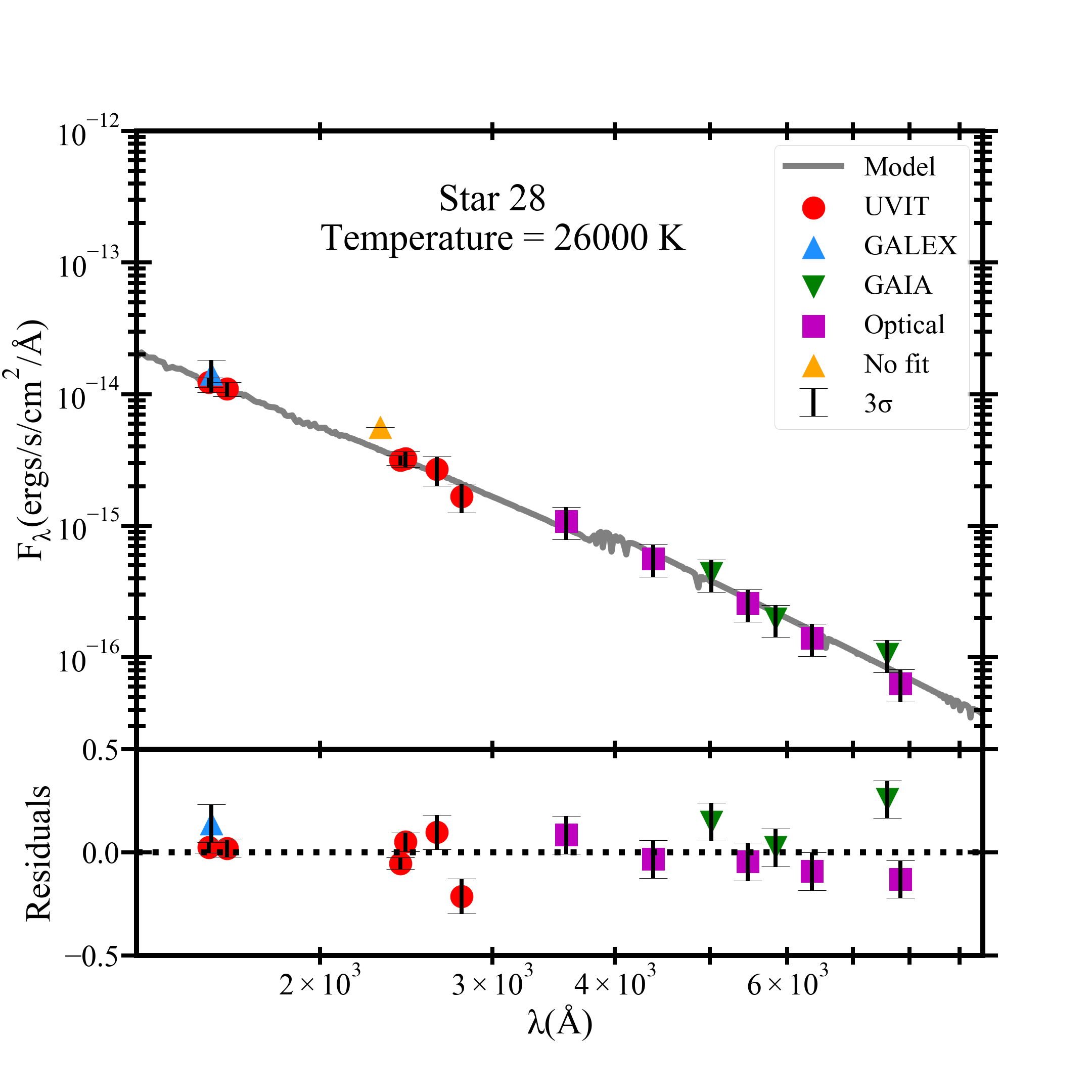}}
}
        \caption{The SEDs for two of the observed UV-bright stars, namely star 3 and star 28, after correcting for extinction. Star 3 lies in the inner region of the cluster and has photometry from the {\it HST} and the UVIT. Star 28 lies in the outer region and has photometric data from UVIT, {\it GALEX}, {\it Gaia} and ground-based optical data from \cite{Stetson2019}. In both the plots, the photometric points excluded from the fitting procedure are shown with orange  symbols. The gray line shows the model spectrum. The residuals of SED fit are shown in the bottom panels of both plots.}
        \label{SED_eg}
\end{figure*}

\begin{figure} [!htb]%
\makebox[0.95\linewidth]
{
    \includegraphics[width=0.55\textwidth]{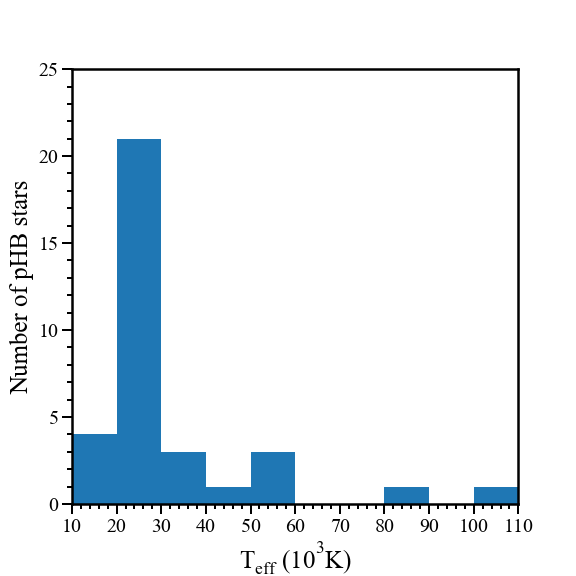}
}
    \caption{Histogram showing the temperature distribution of UV-bright stars in NGC 2808}
    \label{temp_hist}
\end{figure}

Figure \ref{temp_hist} shows a histogram of the temperature distribution of the UV-bright stars in the cluster. The values take a range from 12500 to 100000 K with maximum number of stars having a temperature between 20000 to 30000 K. 


\begin{figure*} [!htb]
\makebox[1\textwidth]
{
    \includegraphics[width=1.25\textwidth]{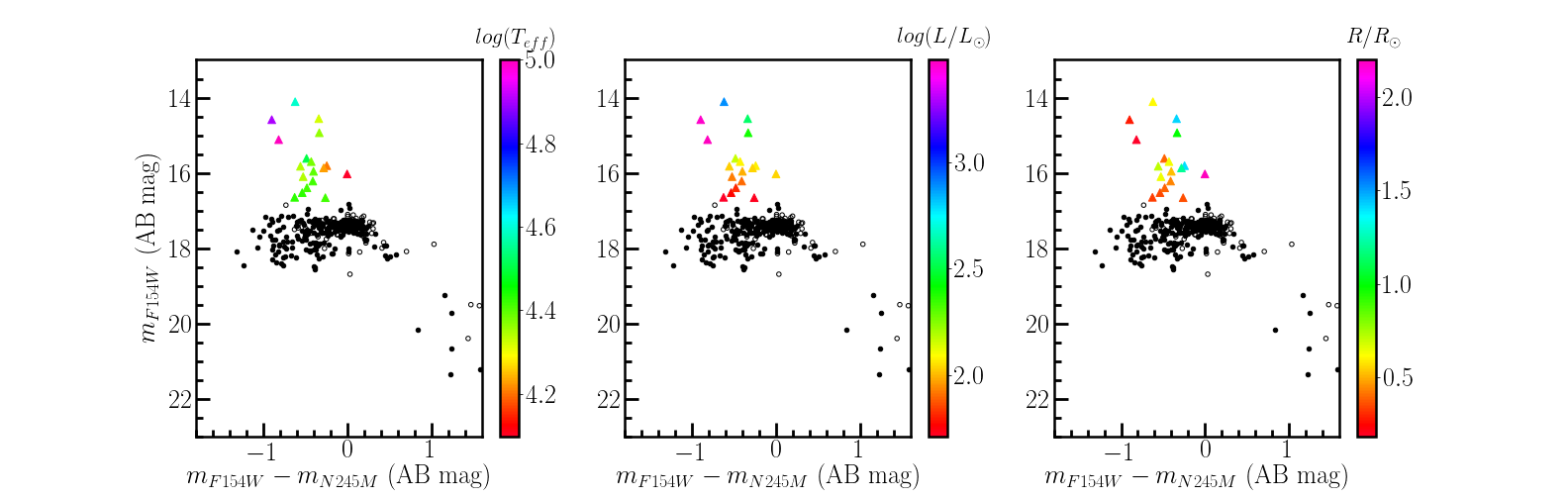}
}
    \caption{The trends in the best-fit parameters (effective temperature, bolometric luminosity and radius) of the 18 UV-bright stars having UVIT photometry shown in the ${\it m}_{F154W}-{\it m}_{N245M}$ vs ${\it m}_{F154W}$ CMD. }
    \label{pHB_colorbars}
\end{figure*}

Figure \ref{pHB_colorbars} shows the trends in the best-fit parameters of the 18 UV-bright stars with UVIT photometry in the ${\it m}_{F154W}-{\it m}_{N245M}$ vs ${\it m}_{F154W}$ CMD. The left, middle and right panels display the trends in T$_{eff}$, luminosity and radius, respectively.  Among the 5 stars with the brightest ${\it m}_{F154W}$ magnitudes, the hottest stars are also the most luminous and have the smallest radii. The coolest among these 5 stars have $log{\it (L/L_{\odot})}$ $\sim$ 2.5 and $R/R_{\odot}$ $\sim$ 1.0. The star with the brightest ${\it m}_{F154W}$ magnitude has log ${\it T_{eff}}$ $\sim$ 4.6, with a fairly high luminosity ($log{\it (L/L_{\odot})}$ $\sim$ 3) and a smaller radius ($R/R_{\odot}$ $\sim$ 0.5). Most of the remaining pHB stars with ${\it m}_{F154W}$ fainter than 15.5 mag have log ${\it T_{eff}}$ in the range 4.3 to 4.5, with $log{\it (L/L_{\odot})}$ $\lesssim$ 2.2 and radii $R/R_{\odot}$ $\lesssim$ 0.5. The three coolest stars with ${\it m}_{F154W}$ $\sim$ 16 mag have $log{\it (L/L_{\odot})}$ $\sim$ 2.0 and are relatively larger in size with $R/R_{\odot}$ $\gtrsim$ 1.5.

In order to evaluate the importance of FUV data points in the estimation of fundamental parameters, we fitted the SEDs of Stars 1 to 15 and Stars 28 to 34 excluding their available FUV photometric points. We found that the parameter values changed significantly for stars hotter than $T_{eff}$ = 40,000 K with the underestimation of $T_{eff}$, $L$ and the overestimation of $R$. This implied that FUV data is crucial to estimate the parameters of very hot stars. Hence, the results derived for hot stars without FUV data points are to be considered with lesser weightage.

Since several previous studies \citep{Bedin2000,Marino2017} adopted a reddening value of $E(B-V)$ = 0.19 mag for this cluster, we also fitted the SEDs using this value to see by what amount the estimated parameter values change. With this reddening measure, we found that the effective temperature values decreased by 10 (eg., in the case of the hottest star) to 20\%, the bolometric luminosities decreased by about 20 to 30\% and the radii increased by about 5 to 10\%.

\section{Evolutionary status of UV-bright stars} \label{evolution tracks}

The luminosities and temperatures derived from SED analysis were used to assess the evolutionary status of the observed UV-bright stars using the Hertzsprung-Russell (HR) diagram. We used the pHB evolutionary tracks from \citet{Moehler2019} which are extended versions of the pAGB evolutionary models developed by \citet{MillerBertolami2016}. The tracks correspond to [M/H] = $-$1, and zero-age main-sequence (ZAMS) mass of ${\it M_{ZAMS}}$ = $0.85$ ${\it M_{\odot}}$ (age = 12 Gyr) assuming scaled-solar metallicity with initial abundances ${\it Z_{ZAMS}}$ = 0.00172, ${\it Y_{ZAMS}}$ = 0.24844, and ${\it X_{ZAMS}}$ = 0.74984. In the models, the RHB, BHB and EHB sequences were populated by regulating the mass loss on the RGB phase. Further details of the evolutionary tracks are presented in the appendix (Table \ref{tab:evol model params}). 

\begin{figure*}[!htb]
\makebox[\textwidth]
{
    \includegraphics[width=1.2\textwidth]{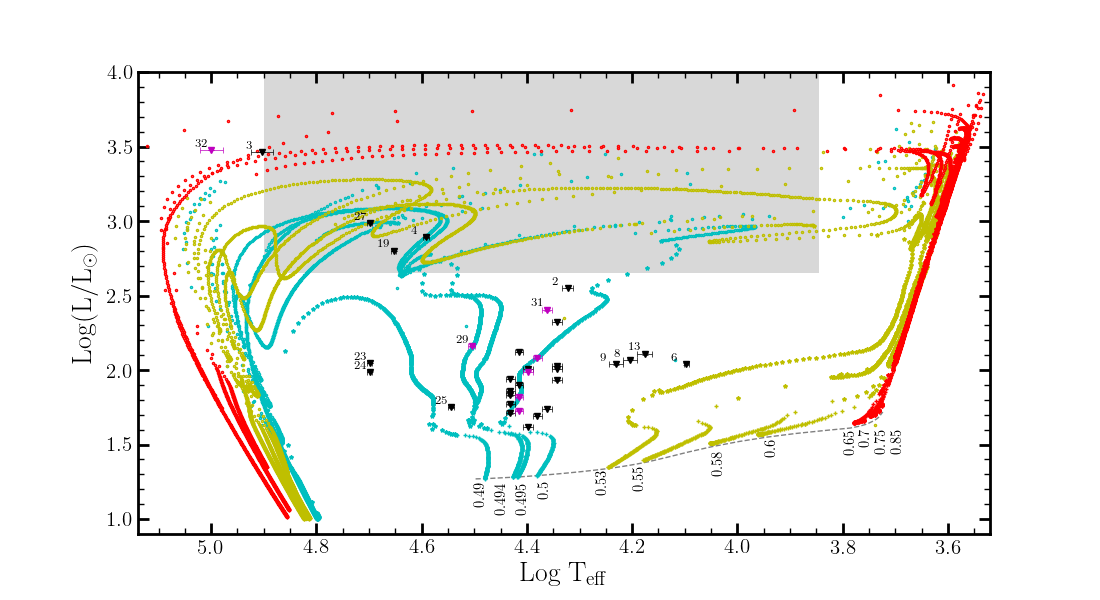}
}
    \caption{Stellar evolutionary tracks starting from ZAHB \citep{Moehler2019}. Red, olive green and cyan colors represent sequences evolving from the RHB, BHB and EHB, respectively. The $+$ symbol indicates HB location and is plotted after every 1 Myr. The stars and circles represent pHB evolution and are plotted with time steps of 0.1 Myr and 1 kyr, respectively.  The value indicated below each sequence represents the mass at the ZAHB for the particular sequence, ${\it M_{ZAHB}}$, in units of ${\it M_{\odot}}$. The gray region marks the domain of hot p(e)AGB stars. The gray dashed line represents the BaSTI ZAHB model. The observed UV-bright stars are plotted with black (located within the inner region of the cluster) and magenta (outer region) symbols with error bars. The numbers marked close to some of the sparsely located stars indicate their IDs from Table \ref{tab:sed params}.}
    \label{pHB_models}
\end{figure*}

Figure \ref{pHB_models} shows the observed UV-bright stars in black (inner region) and magenta (outer region) inverted triangles plotted over the evolutionary tracks. It is evident from this figure that all the UV-bright stars except the three most luminous ones, have evolutionary masses $<$ 0.53 ${\it M_{\odot}}$. Most of these stars are observed to lie along/near the sequence with ${\it M_{ZAHB}}$ = 0.5 ${\it M_{\odot}}$, which is a track evolving from the EHB phase. These are likely to be AGB-manqu\'e stars, which directly descend the white dwarf cooling sequence after core-He exhaustion. The hottest and the most luminous UV-bright star (Star 32 in Table \ref{tab:sed params}) is located between the tracks with ${\it M_{ZAHB}}$ = 0.75 ${\it M_{\odot}}$ and 0.85 ${\it M_{\odot}}$. Star 3, with slightly lower luminosity and ${\it T_{eff}}$, lies at the intersection of two post-RHB sequences with ${\it M_{ZAHB}}$ = 0.65 ${\it M_{\odot}}$ and 0.75 ${\it M_{\odot}}$. From the models, the mass of this star then ranges from 0.527 ${\it M_{\odot}}$ to 0.544 ${\it M_{\odot}}$. Two stars (Star 23 and Star 24) are found to be located outside the range of the model tracks.


\cite{Moehler2019} considered the area defined by  $ log {\it (L/L_{\odot})} > 2.65$ and $4.9 > $ log ${\it T_{eff}} > 3.845$ in the HR diagram (region shaded in gray in Figure \ref{pHB_models}) to be corresponding to hot UV-bright p(e)AGB stars. Going by their definition, from Figure \ref{pHB_models}, we observe three stars to lie inside this region and one star on the high temperature boundary. Among these three stars which are inside the gray-shaded region, two (Star 4 and Star 19) are quite unlikely to be p(e)AGB stars as they are located close to the evolutionary track with ${\it M_{ZAHB}}$ = 0.494 ${\it M_{\odot}}$. The remaining one star (Star 27) occupies a position where post-EHB (cyan) and post-BHB (olive green) sequences overlap, making it difficult to distinguish. Hence, only the hot and the luminous star (Star 3) seen on the boundary of the shaded region can be confirmed as a hot p(e)AGB star. In other words, the cluster has a maximum of 2 and a minimum of 1 detected hot p(e)AGB candidate as per the above definition. We have detected one star (star 32) hotter than the hot boundary of the shaded region. This star might have crossed the p(e)AGB stage and is probably evolving towards the white-dwarf stage.

\subsection{Comparison of theoretically expected and observed numbers of hot p(e)AGB stars}

We now make a comparison of the observed number of hot p(e)AGB stars with the number of such stars anticipated from theoretical predictions. For this, we make use of the evolutionary flux method described in detail in \citet{GreggioRenzini2011}. The following suppositions are made in this calculation : a) GCs are simple stellar populations (SSPs), b) the rate at which stars end up as remnants is equal to the rate at which stars leave the main sequence, c) the life-times of later stages of stellar evolution are considerably shorter than that of main-sequence evolution. Under this premise, the number of stars, ${\it N_{i}}$, in a particular stage of evolution, {\it i}, in a SSP is given by the relation,

\begin{center}
\begin{equation} \label{eq:1}
 {\it   N_{i} = B(t) \times L_{total} \times t_{i}}
\end{equation}
\end{center}

Here, {\it B(t)} is the specific evolutionary flux, which is the number of stars entering (or exiting) a certain phase of stellar evolution per year per luminosity ($L_{\odot}$) of the sampled population; ${\it L_{total}}$ is the total luminosity of the stellar population and ${\it t_{i}}$ is the duration of the evolutionary phase being analyzed. Assuming an age $\sim$ 10 Gyr for the cluster and Salpeter's initial mass function (IMF), we can obtain an approximate value for specific evolutionary flux as ${\it B} \simeq 2 \times 10^{-11}$ stars per year per $L_{\odot}$. 

In this analysis, we consider stars in the hot p(e)AGB phase of evolution, which corresponds to the gray shaded region in Figure \ref{pHB_models}. Each evolutionary track spans a duration, $t$, in this region. In order to estimate the expected number of hot p(e)AGB stars from Eq. \ref{eq:1}, we use the parameters for NGC 2808 listed in Table \ref{ngc2808params}. The anticipated number of p(e)AGB stars in a cluster depends on how frequently a particular evolutionary track is followed in that cluster, which in turn depends on the fraction of stars in various parts of the HB. We calculate the number of p(e)AGB stars evolving along all the evolutionary tracks starting from each branch of HB, i.e., $N^{RHB}$, $N^{BHB}$ \& $N^{EHB}$ for stars evolving from RHB, BHB and EHB respectively, using Eq. \ref{eq:1} \citep{Moehler2019}. For this, we derive the fraction of RHB, BHB and EHB  stars with respect to the total number of HB stars in the cluster (${\it f_{RHB}}$, ${\it f_{BHB}}$ and ${\it f_{EHB}}$) by combining the members from the HUGS and {\it Gaia} catalogs for complete spatial coverage. The values thus obtained are ${\it f_{RHB}}$ = 0.47, ${\it f_{BHB}}$ = 0.35 and ${\it f_{EHB}}$ = 0.07. The final expected number of hot p(e)AGB stars in the cluster is then, 

\small
    \begin{equation}
     {\it   N = (f_{RHB} . N^{RHB}) + (f_{BHB} . N^{BHB}) + (f_{EHB} . N^{EHB})}
    \end{equation}
\normalsize

We obtain a range for ${\it N}$ since ${\it N^{RHB}}$, ${\it N^{BHB}}$ and ${\it N^{EHB}}$ take range of values depending on the duration in the p(e)AGB phase along each evolutionary track. Thus, for NGC 2808, we estimate the expected number of hot p(e)AGB in the gray shaded region of HR diagram as 1.27 to 3.80. The maximum observed number of p(e)AGB stars detected in this shaded region is 2, which is in good agreement with the expected number of such stars.


\begin{table}
\caption{Parameters for NGC 2808}
\makebox[0.86\linewidth]
{
\begin{tabular}{lcr}
\toprule
Parameter & Value & Source \\
\hline
[Fe/H] & $-$1.14 dex & \harris\\
Age    & 10.9 Gyr & \citet{Massari2016} \\
Integrated {\it V} magnitude, ${\it V_{t}}$ & 6.20 mag & \harris\\
Distance modulus, ${\it (m-M)_{V}}$ & 15.59 mag & \harris \\
Reddening, {\it E(B$-$V)} & 0.22 mag & \harris\\
Bolometric correction, ${\it BC_{V}}$ & $-$0.45 mag & \citet{Worthey1994}\\
\toprule
\end{tabular}
}
\label{ngc2808params}
\end{table}

\section{Discussion} \label{discussion}

We identify and characterize the sample of UV-bright member stars in NGC 2808 using the photometric data from the UVIT, in combination with {\it HST}, {\it Gaia} and ground observations. The detection of these stars is extremely important to build a statistically significant sample which can aid in understanding the late phases of evolution of low-mass stars and determining their contribution to the UV luminosity of old stellar systems. As highlighted earlier, observations in the UV wave bands are critical for this purpose. The UV$-$optical and UV CMDs constructed by combining the UVIT data with the {\it HST} (inner $\ang[angle-symbol-over-decimal]{;2.7;} \times \ang[angle-symbol-over-decimal]{;2.7;}$ region of the GC), {\it Gaia} and ground-based optical data (outside {\it HST} FOV) reveal the sequences of hot members in the cluster. We do not use the UVIT data for stars within the innermost $1\arcmin$ diameter region because of possible effects of crowding. In the FUV filters, we detect mainly the pHB, BHB, EHB, BHk and BSS populations, whereas in the NUV filters, the cooler RHB stars can also be spotted additionally. The identification of HB location is supported by the overlaid ZAHB and TAHB models.

\citet{Brown2001} observed the central $\simeq 1750$ $arcsec^{2}$ region of NGC 2808 using {\it HST} STIS FUV/F25QTZ and NUV/F25CN270 filters.  Although a direct comparison cannot be made between their and our UV CMDs (due to the differences in the filter characteristics and the FOV covered), some general features of the CMDs can be examined in detail. In their FUV$-$NUV vs. FUV CMD, a gap between BHB and EHB was found at FUV$-$NUV color $\sim -1$ mag. They also found a sparse sub-luminous population of stars below the end of canonical ZAHB  and explained them by invoking the late-hot flasher (BHk) scenario having enhanced He and C abundances due to flash mixing. The spread in the magnitudes of these stars was accounted for by the evolution of BHk stars to higher luminosities as the He-core burning progressed. We find a similar spread in the BHk magnitudes in our $m_{F154W}-m_{N245M}$ vs $m_{F154W}$ CMD shown in Figure \ref{baf2_b13_cmds}. \citet{Brown2001,Brown2010} noted that some of these BHk stars are unusually redder, i.e., redder than expected from the models for normal BHk or EHB stars. A similar dispersion in the color of BHk stars is observed in our FUV$-$NUV vs. FUV CMD. \citet{Brown2012} analyzed the UV spectra of two BHk stars with redder FUV$-$NUV colors and found that there is large enhancement of Fe-peak elements in these stars which could serve as an explanation for their observed colors. 

\citet{Brown2012} also discussed the results from UV spectral analysis of 3 unclassified objects, U1, U2 and U3, which are hotter than their canonical HB model. Based on their spectra and locations in the UV CMD, the possibilities of them being evolving pAGB stars or white dwarfs were disregarded. U1 and U2 were found to have implausible effective temperatures ($\sim$ 250,000 K) and U3 to have ${\it T_{eff}}$ = 50,000 K. These objects did not have any X-ray counterparts. The authors were unable to conclusively explain these unusually hot objects although propositions such as presence of accretion disk in the objects, change in extinction along the line of sight to the cluster, non-stellar source etc. were put forth. These objects, located within $30\arcsec$ from the cluster center, are not resolved in the UVIT images. However, two of these objects, namely, U1 and U3, could be cross-matched with stars in the HUGS catalog having membership probability $>$ $90\%$. These are found in the BHk and EHB positions respectively in the ${\it m}_{F275W}-{\it m}_{F438W}$ vs. ${\it m}_{F336W}$ CMD shown in Figure \ref{selection_of_inner_stars}. We detect similar objects in our UV CMDs in Figure \ref{baf2_b13_cmds} with bluer ${\it m}_{F154W} - {\it m}_{N245M}$ colors ($\lesssim$ $-$1 mag) than the ZAHB model and roughly the same magnitudes as that of the HB. We plan to characterize the HB stars in our future study.

The study by \cite{Schiavon2012} did not include the full sample of pHB stars in NGC 2808 due to the limited spatial resolution of {\it GALEX} and the lack of membership analysis for these stars. There are 22 candidates in their list of UV-bright stars in NGC 2808. We find 15 stars to be in common between our sample and theirs.  

The UV CMD presented in \citet{Brown2001} shows one pAGB candidate star with FUV magnitude 12.46 mag. This star is unresolved in the UVIT images and not included in the HUGS catalog as it is saturated in the WFC3/UVIS filters. Hence, our analysis does not include this star. \citet{Brown2001} also found 5 pHB candidates from the UV CMD of which 4 stars are common with our catalog. The fifth pHB candidate  is not included in the HUGS catalog, and hence, its membership cannot be assessed even though it is detected in the UVIT images (R.A. = 138.0235 degrees, Decl. = $-$64.86531 degrees).

In Table \ref{teff_compare}, we compare our estimates of $T_{eff}$ with the values available in literature for three UV-bright stars in the cluster. \citet{Brown2012} derived the $T_{eff}$ for the AGB-manqu\'e star AGBM1 using the $HST$ STIS UV spectrum. \citet{Moehler2019} estimated the $T_{eff}$ values for the stars C4594 and C2946 using the medium-resolution optical spectra from the EFOSC2 instrument at the 2.2m MPI/ESO telescope. From the table, it is evident that the $T_{eff}$ values derived through SED analysis are in close agreement with spectroscopic estimations.

\begin{table*} [!htp]
\caption{Table showing the comparison of $T_{eff}$ values derived by us through SED analysis with those available in the literature for 3 UV-bright stars. The errors quoted in our $T_{eff}$ values are estimated as half the model grid step, around the best-fit value, during the SED fit. The first column shows the ID of the stars from Table \ref{tab:sed params}.}
\makebox[0.89\linewidth]
{
\begin{tabular}{ccccc}
\toprule
ID & ID in literature & ${\it T_{eff}}$ from literature &  ${\it T_{eff}}$ from our analysis & Reference \\ 
 &  &   (K)   &  (K)  & \\
\hline
Star 13 & AGBM1 & 14500 & 15500 $\pm$ 500   & \citet{Brown2012} \\
Star 31 & C4594 & 19900 $\pm$ 1600 & 23000 $\pm$ 500 & \citet{Moehler2019}  \\
Star 34 & C2946 & 24900 $\pm$ 1800  & 24000 $\pm$ 500 &  \citet{Moehler2019}\\
\toprule
\end{tabular}%
}
\label{teff_compare}
\end{table*}

\citet{Moehler2019} calculated the theoretically expected numbers of hot p(e)AGB stars in 17 galactic GCs (excluding NGC 2808) and compared them with the observed numbers. The numbers matched more or less in the case of 14 clusters. In the remaining 3 clusters, the observed numbers of p(e)AGB stars were larger than the predicted values. The massive and dense GC, NGC 5139 ($\omega$ Cen) has the maximum number of observed p(e)AGB stars (5 stars), with the expected number in the range 1.3 - 13.5. Here, we find that the number of such stars is between 1 and 3 (including the saturated pAGB candidate from \citet{Brown2001}), which agrees well with the expected number for NGC 2808.


The pHB stars identified in this study will provide a good sample to explore the stellar evolutionary properties in this phase. From Figure \ref{pHB_spatial_location}, we see that there are seven member stars located in the outer region of the cluster that are ideal candidates for further follow-up spectroscopic studies. These are identified from the {\it AstroSat}/UVIT images and are in relatively less crowded regions of the cluster. This GC is known to host populations with a wide range of main-sequence He abundances \citep{Piotto2007,Milone2015}. In addition, many of the UV-bright stars with $T_{eff}$ $>$ 25,000 K are likely to experience gravitational settling, and/or weak winds which may alter their surface helium (and metal) abundance (eg., \citet{Dixon2017}). These abundance uncertainties call for thorough spectroscopic investigation. Moreover, spectroscopy can provide accurate estimation of $T_{eff}$, log $\textit{g}$, radial velocity and its variation (if any), detailed chemical abundance information on the signs of third dredge-up (which will help to clearly distinguish between peAGB and pAGB phases), and so on.


\section{Summary} \label{summary}

1. We performed a comprehensive study of the UV-bright member stars in the GC NGC 2808 using the {\it AstroSat}/UVIT, {\it HST}, {\it Gaia} DR2 and ground-based optical data. The identification and detailed study of these stars are important to create a statistically significant sample for two main reasons : (i) to throw light on the rapid evolution of late phases of low mass stars such as pAGB, peAGB and AGB-manqu\'e phases, (ii) to assess the contribution of these stars to the total UV output of old stellar systems. 

2. Member stars in the inner and outer parts of the cluster are identified to create optical, UV-optical and UV CMDs. The stars in the HB sequence are identified and compared with the ZAHB and TAHB models. A large number of hot HB stars are detected along with a FUV-bright BSS. These will be studied in detail in the future.

3. We detected 34 UV-bright stars based on their locations in the UV CMDs. Among these, 27 stars are found to be located within the inner $\ang[angle-symbol-over-decimal]{;2.7;} \times \ang[angle-symbol-over-decimal]{;2.7;}$ region of the cluster and 7 stars in the outer region. 

4. We estimated parameters such as ${\it T_{eff}}$, $R/R_{\odot}$ and $L/L_{\odot}$ of these stars through SED fitting technique. Their effective temperatures range from 12500 K to 100,000 K, luminosities from $ \sim 40$ to $3000$ $L_{\odot}$ and radii from  0.13 to 2.2 $R_{\odot}$. Our ${\it T_{eff}}$ estimations from SED fitting are found to match well with the available spectroscopic estimations for a few stars from literature. 

5. By comparing the derived parameters with theoretical models available for evolved stellar populations, the evolutionary status of these stars is probed. We find that most UV-bright stars have evolved from EHB stars with ${\it {\it M_{ZAHB}}}$ = 0.5 ${\it M_{\odot}}$, and these are in the AGB-manqu\'e phase. From the theoretical models, we observe that all except the three hottest and the most luminous UV-bright stars have HB progenitors with ${\it M_{ZAHB}}$ $<$ 0.53 ${\it M_{\odot}}$. 

6. The expected number of hot p(e)AGB stars in NGC 2808 is estimated from stellar evolutionary models and is found to agree well with the observed number. 
 
7. Seven pHB stars identified in the outer region are ideal for further spectroscopic follow-up studies. These stars are identified from the {\it AstroSat}/UVIT images. This work thus demonstrates the capability of the UVIT in detecting and characterizing the UV-bright stars. 


\acknowledgments

We thank the anonymous referee for the encouraging comments and suggestions. We are thankful to S. Moehler for her valuable comments which helped in improving the manuscript. We thank Gaurav Singh for providing us the catalog of {\it Gaia} proper-motion based cluster members. DSP thanks Sharmila Rani and Vikrant Jadhav for useful discussions. This publication utilizes the data from {\it AstroSat} mission's UVIT, which is archived at the Indian Space Science Data Centre (ISSDC). The UVIT project is a result of collaboration between IIA, Bengaluru, IUCAA, Pune, TIFR, Mumbai, several centers of ISRO, and
CSA. This research made use of VOSA, developed under
the Spanish Virtual Observatory project supported by the
Spanish MINECO through grant AyA2017-84089. This research also made use of Topcat \citep{Taylor2005}, Aladin sky atlas developed at CDS, Strasbourg Observatory, France \citep{Bonnarel2000,Boch2014}, Matplotlib \citep{Hunter2007}, NumPy \citep{van2011}, SciPy \citep{Virtanen_2020}, pandas \citep{McKinney_2010, McKinney_2011}. 


\bibliography{ngc2808}{}

\newcommand{\noop}[0]{}
\begin{thebibliography}{}
\expandafter\ifx\csname natexlab\endcsname\relax\def\natexlab#1{#1}\fi
\providecommand{\url}[1]{\href{#1}{#1}}
\providecommand{\dodoi}[1]{doi:~\href{http://doi.org/#1}{\nolinkurl{#1}}}
\providecommand{\doeprint}[1]{\href{http://ascl.net/#1}{\nolinkurl{http://ascl.net/#1}}}
\providecommand{\doarXiv}[1]{\href{https://arxiv.org/abs/#1}{\nolinkurl{https://arxiv.org/abs/#1}}}

\bibitem[{{Bayo} {et~al.}(2008){Bayo}, {Rodrigo}, {Barrado Y Navascu{\'e}s},
  {Solano}, {Guti{\'e}rrez}, {Morales-Calder{\'o}n}, \& {Allard}}]{Bayo2008}
{Bayo}, A., {Rodrigo}, C., {Barrado Y Navascu{\'e}s}, D., {et~al.} 2008, \aap,
  492, 277

\bibitem[{{Bedin} {et~al.}(2000){Bedin}, {Piotto}, {Zoccali}, {Stetson},
  {Saviane}, {Cassisi}, \& {Bono}}]{Bedin2000}
{Bedin}, L.~R., {Piotto}, G., {Zoccali}, M., {et~al.} 2000, \aap, 363, 159

\bibitem[{{Boch} \& {Fernique}(2014)}]{Boch2014}
{Boch}, T., \& {Fernique}, P. 2014, in Astronomical Society of the Pacific
  Conference Series, Vol. 485, Astronomical Data Analysis Software and Systems
  XXIII, ed. N.~{Manset} \& P.~{Forshay}, 277

\bibitem[{{Bonnarel} {et~al.}(2000){Bonnarel}, {Fernique}, {Bienaym{\'e}},
  {Egret}, {Genova}, {Louys}, {Ochsenbein}, {Wenger}, \&
  {Bartlett}}]{Bonnarel2000}
{Bonnarel}, F., {Fernique}, P., {Bienaym{\'e}}, O., {et~al.} 2000, \aaps, 143,
  33

\bibitem[{{Brocato} {et~al.}(1990){Brocato}, {Matteucci}, {Mazzitelli}, \&
  {Tornambe}}]{Brocato1990}
{Brocato}, E., {Matteucci}, F., {Mazzitelli}, I., \& {Tornambe}, A. 1990, \apj,
  349, 458

\bibitem[{{Brown} {et~al.}(2000){Brown}, {Bowers}, {Kimble}, {Sweigart}, \&
  {Ferguson}}]{Brown2000}
{Brown}, T.~M., {Bowers}, C.~W., {Kimble}, R.~A., {Sweigart}, A.~V., \&
  {Ferguson}, H.~C. 2000, \apj, 532, 308

\bibitem[{{Brown} {et~al.}(1997){Brown}, {Ferguson}, {Davidsen}, \&
  {Dorman}}]{Brown1997}
{Brown}, T.~M., {Ferguson}, H.~C., {Davidsen}, A.~F., \& {Dorman}, B. 1997,
  \apj, 482, 685

\bibitem[{{Brown} {et~al.}(2012){Brown}, {Lanz}, {Sweigart}, {Cracraft},
  {Hubeny}, \& {Landsman}}]{Brown2012}
{Brown}, T.~M., {Lanz}, T., {Sweigart}, A.~V., {et~al.} 2012, \apj, 748, 85

\bibitem[{{Brown} {et~al.}(2001){Brown}, {Sweigart}, {Lanz}, {Land sman}, \&
  {Hubeny}}]{Brown2001}
{Brown}, T.~M., {Sweigart}, A.~V., {Lanz}, T., {Land sman}, W.~B., \& {Hubeny},
  I. 2001, \apj, 562, 368

\bibitem[{{Brown} {et~al.}(2010){Brown}, {Sweigart}, {Lanz}, {Smith},
  {Landsman}, \& {Hubeny}}]{Brown2010}
{Brown}, T.~M., {Sweigart}, A.~V., {Lanz}, T., {et~al.} 2010, \apj, 718, 1332

\bibitem[{{Brown} {et~al.}(2016){Brown}, {Cassisi}, {D'Antona}, {Salaris},
  {Milone}, {Dalessandro}, {Piotto}, {Renzini}, {Sweigart}, {Bellini},
  {Ortolani}, {Sarajedini}, {Aparicio}, {Bedin}, {Anderson}, {Pietrinferni}, \&
  {Nardiello}}]{Brown2016}
{Brown}, T.~M., {Cassisi}, S., {D'Antona}, F., {et~al.} 2016, \apj, 822, 44

\bibitem[{{Cardelli} {et~al.}(1989){Cardelli}, {Clayton}, \&
  {Mathis}}]{Cardelli1989}
{Cardelli}, J.~A., {Clayton}, G.~C., \& {Mathis}, J.~S. 1989, \apj, 345, 245

\bibitem[{{Castelli} {et~al.}(1997){Castelli}, {Gratton}, \&
  {Kurucz}}]{Castelli1997}
{Castelli}, F., {Gratton}, R.~G., \& {Kurucz}, R.~L. 1997, \aap, 318, 841

\bibitem[{{Castelli} \& {Kurucz}(2003)}]{Castelli2003}
{Castelli}, F., \& {Kurucz}, R.~L. 2003, in IAU Symposium, Vol. 210, Modelling
  of Stellar Atmospheres, ed. N.~{Piskunov}, W.~W. {Weiss}, \& D.~F. {Gray},
  A20

\bibitem[{{Charpinet} {et~al.}(2011){Charpinet}, {Van Grootel}, {Fontaine},
  {Green}, {Brassard}, {Randall}, {Silvotti}, {{\O}stensen}, {Kjeldsen},
  {Christensen-Dalsgaard}, {Kawaler}, {Clarke}, {Li}, \&
  {Wohler}}]{Charpinet2011}
{Charpinet}, S., {Van Grootel}, V., {Fontaine}, G., {et~al.} 2011, \aap, 530,
  A3

\bibitem[{{Chayer} {et~al.}(2015){Chayer}, {Dixon}, {Fullerton},
  {Ooghe-Tabanou}, \& {Reid}}]{Chayar2015}
{Chayer}, P., {Dixon}, W.~V., {Fullerton}, A.~W., {Ooghe-Tabanou}, B., \&
  {Reid}, I.~N. 2015, \mnras, 452, 2292

\bibitem[{{Constantino} {et~al.}(2015){Constantino}, {Campbell},
  {Christensen-Dalsgaard}, {Lattanzio}, \& {Stello}}]{Constantino2015}
{Constantino}, T., {Campbell}, S.~W., {Christensen-Dalsgaard}, J., {Lattanzio},
  J.~C., \& {Stello}, D. 2015, \mnras, 452, 123

\bibitem[{{Dixon} {et~al.}(2017){Dixon}, {Chayer}, {Latour}, {Miller
  Bertolami}, \& {Benjamin}}]{Dixon2017}
{Dixon}, W.~V., {Chayer}, P., {Latour}, M., {Miller Bertolami}, M.~M., \&
  {Benjamin}, R.~A. 2017, \aj, 154, 126

\bibitem[{{Dixon} {et~al.}(2019){Dixon}, {Chayer}, {Reid}, \& {Miller
  Bertolami}}]{Dixon2019}
{Dixon}, W.~V., {Chayer}, P., {Reid}, I.~N., \& {Miller Bertolami}, M.~M. 2019,
  \aj, 157, 147

\bibitem[{{Dorman} {et~al.}(1995){Dorman}, {O'Connell}, \& {Rood}}]{Dorman1995}
{Dorman}, B., {O'Connell}, R.~W., \& {Rood}, R.~T. 1995, \apj, 442, 105

\bibitem[{{Dorman} {et~al.}(1993){Dorman}, {Rood}, \& {O'Connell}}]{Dorman1993}
{Dorman}, B., {Rood}, R.~T., \& {O'Connell}, R.~W. 1993, \apj, 419, 596

\bibitem[{{Fitzpatrick}(1999)}]{Fitzpatrick1999}
{Fitzpatrick}, E.~L. 1999, \pasp, 111, 63

\bibitem[{{Gaia Collaboration} {et~al.}(2018){Gaia Collaboration}, {Helmi},
  {van Leeuwen}, {McMillan}, {Massari}, {Antoja}, {Robin}, {Lindegren},
  {Bastian}, {Arenou}, {Babusiaux}, {Biermann}, {Breddels}, {Hobbs}, {Jordi},
  {Pancino}, {Reyl{\'e}}, {Veljanoski}, {Brown}, {Vallenari}, {Prusti}, {de
  Bruijne}, {Bailer-Jones}, {Evans}, {Eyer}, {Jansen}, {Klioner}, {Lammers},
  {Luri}, {Mignard}, {Panem}, {Pourbaix}, {Randich}, {Sartoretti}, {Siddiqui},
  {Soubiran}, {Walton}, {Cropper}, {Drimmel}, {Katz}, {Lattanzi}, {Bakker},
  {Cacciari}, {Casta{\~n}eda}, {Chaoul}, {Cheek}, {De Angeli}, {Fabricius},
  {Guerra}, {Holl}, {Masana}, {Messineo}, {Mowlavi}, {Nienartowicz}, {Panuzzo},
  {Portell}, {Riello}, {Seabroke}, {Tanga}, {Th{\'e}venin}, {Gracia-Abril},
  {Comoretto}, {Garcia-Reinaldos}, {Teyssier}, {Altmann}, {Andrae}, {Audard},
  {Bellas-Velidis}, {Benson}, {Berthier}, {Blomme}, {Burgess}, {Busso},
  {Carry}, {Cellino}, {Clementini}, {Clotet}, {Creevey}, {Davidson}, {De
  Ridder}, {Delchambre}, {Dell'Oro}, {Ducourant},
  {Fern{\'a}ndez-Hern{\'a}ndez}, {Fouesneau}, {Fr{\'e}mat}, {Galluccio},
  {Garc{\'\i}a-Torres}, {Gonz{\'a}lez-N{\'u}{\~n}ez}, {Gonz{\'a}lez-Vidal},
  {Gosset}, {Guy}, {Halbwachs}, {Hambly}, {Harrison}, {Hern{\'a}ndez},
  {Hestroffer}, {Hodgkin}, {Hutton}, {Jasniewicz}, {Jean-Antoine-Piccolo},
  {Jordan}, {Korn}, {Krone-Martins}, {Lanzafame}, {Lebzelter}, {L{\"o}ffler},
  {Manteiga}, {Marrese}, {Mart{\'\i}n-Fleitas}, {Moitinho}, {Mora}, {Muinonen},
  {Osinde}, {Pauwels}, {Petit}, {Recio-Blanco}, {Richards}, {Rimoldini},
  {Sarro}, {Siopis}, {Smith}, {Sozzetti}, {S{\"u}veges}, {Torra}, {van Reeven},
  {Abbas}, {Abreu Aramburu}, {Accart}, {Aerts}, {Altavilla}, {{\'A}lvarez},
  {Alvarez}, {Alves}, {Anderson}, {Andrei}, {Anglada Varela}, {Antiche},
  {Arcay}, {Astraatmadja}, {Bach}, {Baker}, {Balaguer-N{\'u}{\~n}ez}, {Balm},
  {Barache}, {Barata}, {Barbato}, {Barblan}, {Barklem}, {Barrado}, {Barros},
  {Barstow}, {Bartholom{\'e} Mu{\~n}oz}, {Bassilana}, {Becciani}, {Bellazzini},
  {Berihuete}, {Bertone}, {Bianchi}, {Bienaym{\'e}}, {Blanco-Cuaresma}, {Boch},
  {Boeche}, {Bombrun}, {Borrachero}, {Bossini}, {Bouquillon}, {Bourda},
  {Bragaglia}, {Bramante}, {Bressan}, {Brouillet}, {Br{\"u}semeister},
  {Brugaletta}, {Bucciarelli}, {Burlacu}, {Busonero}, {Butkevich}, {Buzzi},
  {Caffau}, {Cancelliere}, {Cannizzaro}, {Cantat-Gaudin}, {Carballo},
  {Carlucci}, {Carrasco}, {Casamiquela}, {Castellani}, {Castro-Ginard},
  {Charlot}, {Chemin}, {Chiavassa}, {Cocozza}, {Costigan}, {Cowell}, {Crifo},
  {Crosta}, {Crowley}, {Cuypers}, {Dafonte}, {Damerdji}, {Dapergolas}, {David},
  {David}, {de Laverny}, {De Luise}, {De March}, {de Martino}, {de Souza}, {de
  Torres}, {Debosscher}, {del Pozo}, {Delbo}, {Delgado}, {Delgado}, {Di
  Matteo}, {Diakite}, {Diener}, {Distefano}, {Dolding}, {Drazinos},
  {Dur{\'a}n}, {Edvardsson}, {Enke}, {Eriksson}, {Esquej}, {Eynard Bontemps},
  {Fabre}, {Fabrizio}, {Faigler}, {Falc{\~a}o}, {Farr{\`a}s Casas}, {Federici},
  {Fedorets}, {Fernique}, {Figueras}, {Filippi}, {Findeisen}, {Fonti},
  {Fraile}, {Fraser}, {Fr{\'e}zouls}, {Gai}, {Galleti}, {Garabato},
  {Garc{\'\i}a-Sedano}, {Garofalo}, {Garralda}, {Gavel}, {Gavras}, {Gerssen},
  {Geyer}, {Giacobbe}, {Gilmore}, {Girona}, {Giuffrida}, {Glass}, {Gomes},
  {Granvik}, {Gueguen}, {Guerrier}, {Guiraud}, {Guti{\'e}rrez-S{\'a}nchez},
  {Hofmann}, {Holland}, {Huckle}, {Hypki}, {Icardi}, {Jan{\ss}en}, {Jevardat de
  Fombelle}, {Jonker}, {Juh{\'a}sz}, {Julbe}, {Karampelas}, {Kewley}, {Klar},
  {Kochoska}, {Kohley}, {Kolenberg}, {Kontizas}, {Kontizas}, {Koposov},
  {Kordopatis}, {Kostrzewa-Rutkowska}, {Koubsky}, {Lambert}, {Lanza}, {Lasne},
  {Lavigne}, {Le Fustec}, {Le Poncin-Lafitte}, {Lebreton}, {Leccia}, {Leclerc},
  {Lecoeur-Taibi}, {Lenhardt}, {Leroux}, {Liao}, {Licata}, {Lindstr{\o}m},
  {Lister}, {Livanou}, {Lobel}, {L{\'o}pez}, {Managau}, {Mann}, {Mantelet},
  {Marchal}, {Marchant}, {Marconi}, {Marinoni}, {Marschalk{\'o}}, {Marshall},
  {Martino}, {Marton}, {Mary}, {Matijevi{\v{c}}}, {Mazeh}, {Messina},
  {Michalik}, {Millar}, {Molina}, {Molinaro}, {Moln{\'a}r}, {Montegriffo},
  {Mor}, {Morbidelli}, {Morel}, {Morris}, {Mulone}, {Muraveva}, {Musella},
  {Nelemans}, {Nicastro}, {Noval}, {O'Mullane}, {Ord{\'e}novic},
  {Ord{\'o}{\~n}ez-Blanco}, {Osborne}, {Pagani}, {Pagano}, {Pailler},
  {Palacin}, {Palaversa}, {Panahi}, {Pawlak}, {Piersimoni}, {Pineau}, {Plachy},
  {Plum}, {Poggio}, {Poujoulet}, {Pr{\v{s}}a}, {Pulone}, {Racero}, {Ragaini},
  {Rambaux}, {Ramos-Lerate}, {Regibo}, {Riclet}, {Ripepi}, {Riva}, {Rivard},
  {Rixon}, {Roegiers}, {Roelens}, {Romero-G{\'o}mez}, {Rowell}, {Royer},
  {Ruiz-Dern}, {Sadowski}, {Sagrist{\`a} Sell{\'e}s}, {Sahlmann}, {Salgado},
  {Salguero}, {Sanna}, {Santana-Ros}, {Sarasso}, {Savietto}, {Schultheis},
  {Sciacca}, {Segol}, {Segovia}, {S{\'e}gransan}, {Shih}, {Siltala}, {Silva},
  {Smart}, {Smith}, {Solano}, {Solitro}, {Sordo}, {Soria Nieto}, {Souchay},
  {Spagna}, {Spoto}, {Stampa}, {Steele}, {Steidelm{\"u}ller}, {Stephenson},
  {Stoev}, {Suess}, {Surdej}, {Szabados}, {Szegedi-Elek}, {Tapiador}, {Taris},
  {Tauran}, {Taylor}, {Teixeira}, {Terrett}, {Teyssand ier}, {Thuillot},
  {Titarenko}, {Torra Clotet}, {Turon}, {Ulla}, {Utrilla}, {Uzzi}, {Vaillant},
  {Valentini}, {Valette}, {van Elteren}, {Van Hemelryck}, {van Leeuwen},
  {Vaschetto}, {Vecchiato}, {Viala}, {Vicente}, {Vogt}, {von Essen}, {Voss},
  {Votruba}, {Voutsinas}, {Walmsley}, {Weiler}, {Wertz}, {Wevems},
  {Wyrzykowski}, {Yoldas}, {{\v{Z}}erjal}, {Ziaeepour}, {Zorec}, {Zschocke},
  {Zucker}, {Zurbach}, \& {Zwitter}}]{Helmi2018}
{Gaia Collaboration}, {Helmi}, A., {van Leeuwen}, F., {et~al.} 2018, \aap, 616,
  A12

\bibitem[{{Greggio} \& {Renzini}(1990)}]{GreggioRenzini1990}
{Greggio}, L., \& {Renzini}, A. 1990, \apj, 364, 35

\bibitem[{{Greggio} \& {Renzini}(1999)}]{GreggioRenzini1999}
---. 1999, \memsai, 70, 691

\bibitem[{{Greggio} \& {Renzini}(2011)}]{GreggioRenzini2011}
---. 2011, The Fundamentals of Evolutionary Population Synthesis (John Wiley \&
  Sons, Ltd), 35--59

\bibitem[{{Harris}(1996)}]{Harris1996}
{Harris}, W.~E. 1996, \aj, 112, 1487

\bibitem[{{Harris}(2018)}]{WEHarris2018}
---. 2018, \aj, 156, 296

\bibitem[{{Hidalgo} {et~al.}(2018){Hidalgo}, {Pietrinferni}, {Cassisi},
  {Salaris}, {Mucciarelli}, {Savino}, {Aparicio}, {Silva Aguirre}, \&
  {Verma}}]{Hidalgo2018}
{Hidalgo}, S.~L., {Pietrinferni}, A., {Cassisi}, S., {et~al.} 2018, \apj, 856,
  125

\bibitem[{Hunter(2007)}]{Hunter2007}
Hunter, J.~D. 2007, Computing in Science \& Engineering, 9, 90

\bibitem[{{Jain} {et~al.}(2019){Jain}, {Vig}, \& {Ghosh}}]{Jain2019}
{Jain}, R., {Vig}, S., \& {Ghosh}, S.~K. 2019, \mnras, 485, 2877

\bibitem[{{Kunder} {et~al.}(2013){Kunder}, {Stetson}, {Catelan}, {Walker}, \&
  {Amigo}}]{Kunder2013}
{Kunder}, A., {Stetson}, P.~B., {Catelan}, M., {Walker}, A.~R., \& {Amigo}, P.
  2013, \aj, 145, 33

\bibitem[{{Marino} {et~al.}(2017){Marino}, {Milone}, {Yong}, {Da Costa},
  {Asplund}, {Bedin}, {Jerjen}, {Nardiello}, {Piotto}, {Renzini}, \&
  {Shetrone}}]{Marino2017}
{Marino}, A.~F., {Milone}, A.~P., {Yong}, D., {et~al.} 2017, \apj, 843, 66,
  \dodoi{10.3847/1538-4357/aa7852}

\bibitem[{{Massari} {et~al.}(2016){Massari}, {Fiorentino}, {McConnachie},
  {Bono}, {Dall'Ora}, {Ferraro}, {Iannicola}, {Stetson}, {Turri}, \&
  {Tolstoy}}]{Massari2016}
{Massari}, D., {Fiorentino}, G., {McConnachie}, A., {et~al.} 2016, \aap, 586,
  A51

\bibitem[{{McDonald} \& {Zijlstra}(2015)}]{Mcdonald2015}
{McDonald}, I., \& {Zijlstra}, A.~A. 2015, \mnras, 448, 502

\bibitem[{McKinney(2010)}]{McKinney_2010}
McKinney, W. 2010, in Proceedings of the 9th Python in Science Conference, Vol.
  445, Austin, TX, 51--56

\bibitem[{McKinney(2011)}]{McKinney_2011}
McKinney, W. 2011, Python for High Performance and Scientific Computing, 14

\bibitem[{{Miller Bertolami}(2016)}]{MillerBertolami2016}
{Miller Bertolami}, M.~M. 2016, \aap, 588, A25

\bibitem[{{Milone} {et~al.}(2015){Milone}, {Marino}, {Piotto}, {Renzini},
  {Bedin}, {Anderson}, {Cassisi}, {D'Antona}, {Bellini}, {Jerjen},
  {Pietrinferni}, \& {Ventura}}]{Milone2015}
{Milone}, A.~P., {Marino}, A.~F., {Piotto}, G., {et~al.} 2015, \apj, 808, 51,
  \dodoi{10.1088/0004-637X/808/1/51}

\bibitem[{{Moehler}(2001)}]{Moehler2001}
{Moehler}, S. 2001, \pasp, 113, 1162

\bibitem[{{Moehler}(2010)}]{Moehler2010}
---. 2010, \memsai, 81, 838

\bibitem[{{Moehler} {et~al.}(2019){Moehler}, {Landsman}, {Lanz}, \& {Miller
  Bertolami}}]{Moehler2019}
{Moehler}, S., {Landsman}, W.~B., {Lanz}, T., \& {Miller Bertolami}, M.~M.
  2019, \aap, 627, A34

\bibitem[{{Nardiello} {et~al.}(2018){Nardiello}, {Libralato}, {Piotto},
  {Anderson}, {Bellini}, {Aparicio}, {Bedin}, {Cassisi}, {Granata}, {King},
  {Lucertini}, {Marino}, {Milone}, {Ortolani}, {Platais}, \& {van der
  Marel}}]{Nardiello2018}
{Nardiello}, D., {Libralato}, M., {Piotto}, G., {et~al.} 2018, \mnras, 481,
  3382

\bibitem[{O'Connell(1999)}]{O'Connell1999}
O'Connell, R.~W. 1999, Annual Review of Astronomy and Astrophysics, 37, 603

\bibitem[{{Piotto} {et~al.}(2007){Piotto}, {Bedin}, {Anderson}, {King},
  {Cassisi}, {Milone}, {Villanova}, {Pietrinferni}, \& {Renzini}}]{Piotto2007}
{Piotto}, G., {Bedin}, L.~R., {Anderson}, J., {et~al.} 2007, \apjl, 661, L53,
  \dodoi{10.1086/518503}

\bibitem[{{Piotto} {et~al.}(2015){Piotto}, {Milone}, {Bedin}, {Anderson},
  {King}, {Marino}, {Nardiello}, {Aparicio}, {Barbuy}, {Bellini}, {Brown},
  {Cassisi}, {Cool}, {Cunial}, {Dalessandro}, {D'Antona}, {Ferraro}, {Hidalgo},
  {Lanzoni}, {Monelli}, {Ortolani}, {Renzini}, {Salaris}, {Sarajedini}, {van
  der Marel}, {Vesperini}, \& {Zoccali}}]{Piotto2015}
{Piotto}, G., {Milone}, A.~P., {Bedin}, L.~R., {et~al.} 2015, \aj, 149, 91

\bibitem[{{Postma} \& {Leahy}(2017)}]{Postma2017}
{Postma}, J.~E., \& {Leahy}, D. 2017, \pasp, 129, 115002

\bibitem[{{Raso} {et~al.}(2017){Raso}, {Ferraro}, {Dalessandro}, {Lanzoni},
  {Nardiello}, {Bellini}, \& {Vesperini}}]{Raso2017}
{Raso}, S., {Ferraro}, F.~R., {Dalessandro}, E., {et~al.} 2017, \apj, 839, 64

\bibitem[{{Rauch} \& {Deetjen}(2003)}]{Rauch2003}
{Rauch}, T., \& {Deetjen}, J. 2003, in Astronomical Society of the Pacific
  Conference Series, Vol. 288, Stellar Atmosphere Modeling, ed. I.~{Hubeny},
  D.~{Mihalas}, \& K.~{Werner}, 103

\bibitem[{{Salaris} {et~al.}(2016){Salaris}, {Cassisi}, \&
  {Pietrinferni}}]{Salaris2016}
{Salaris}, M., {Cassisi}, S., \& {Pietrinferni}, A. 2016, \aap, 590, A64

\bibitem[{{Schiavon} {et~al.}(2012){Schiavon}, {Dalessandro}, {Sohn}, {Rood},
  {O'Connell}, {Ferraro}, {Lanzoni}, {Beccari}, {Rey}, {Rhee}, {Rich}, {Yoon},
  \& {Lee}}]{Schiavon2012}
{Schiavon}, R.~P., {Dalessandro}, E., {Sohn}, S.~T., {et~al.} 2012, \aj, 143,
  121

\bibitem[{{Singh} {et~al.}(2020){Singh}, {Sahu}, {Subramaniam}, \&
  {Yadav}}]{Singh2020}
{Singh}, G., {Sahu}, S., {Subramaniam}, A., \& {Yadav}, R.~K.~S. 2020, arXiv
  e-prints, arXiv:2010.06979.
\newblock \doarXiv{2010.06979}

\bibitem[{{Sirianni} {et~al.}(2005){Sirianni}, {Jee}, {Ben{\'\i}tez},
  {Blakeslee}, {Martel}, {Meurer}, {Clampin}, {De Marchi}, {Ford}, {Gilliland},
  {Hartig}, {Illingworth}, {Mack}, \& {McCann}}]{Sirianni2005}
{Sirianni}, M., {Jee}, M.~J., {Ben{\'\i}tez}, N., {et~al.} 2005, \pasp, 117,
  1049, \dodoi{10.1086/444553}

\bibitem[{Stetson(1987)}]{Stetson1987}
Stetson, P.~B. 1987, Publications of the Astronomical Society of the Pacific,
  99, 191

\bibitem[{{Stetson} {et~al.}(2019){Stetson}, {Pancino}, {Zocchi}, {Sanna}, \&
  {Monelli}}]{Stetson2019}
{Stetson}, P.~B., {Pancino}, E., {Zocchi}, A., {Sanna}, N., \& {Monelli}, M.
  2019, \mnras, 485, 3042

\bibitem[{{Tandon} {et~al.}(2017){Tandon}, {Subramaniam}, {Girish}, {Postma},
  {Sankarasubramanian}, {Sriram}, {Stalin}, {Mondal}, {Sahu}, {Joseph},
  {Hutchings}, {Ghosh}, {Barve}, {George}, {Kamath}, {Kathiravan}, {Kumar},
  {Lancelot}, {Leahy}, {Mahesh}, {Mohan}, {Nagabhushana}, {Pati}, {Kameswara
  Rao}, {Sreedhar}, \& {Sreekumar}}]{Tandon2017}
{Tandon}, S.~N., {Subramaniam}, A., {Girish}, V., {et~al.} 2017, \aj, 154, 128

\bibitem[{{Taylor}(2005)}]{Taylor2005}
{Taylor}, M.~B. 2005, in Astronomical Society of the Pacific Conference Series,
  Vol. 347, Astronomical Data Analysis Software and Systems XIV, ed.
  P.~{Shopbell}, M.~{Britton}, \& R.~{Ebert}, 29

\bibitem[{{Thompson} {et~al.}(2007){Thompson}, {Keenan}, {Dufton}, {Ryans},
  {Smoker}, {Lambert}, \& {Zijlstra}}]{Thompson2007}
{Thompson}, H.~M.~A., {Keenan}, F.~P., {Dufton}, P.~L., {et~al.} 2007, \mnras,
  378, 1619

\bibitem[{Van Der~Walt {et~al.}(2011)Van Der~Walt, Colbert, \&
  Varoquaux}]{van2011}
Van Der~Walt, S., Colbert, S.~C., \& Varoquaux, G. 2011, Computing in Science
  \& Engineering, 13, 22

\bibitem[{{Virtanen} {et~al.}(2020){Virtanen}, {Gommers}, {Oliphant},
  {Haberland}, {Reddy}, {Cournapeau}, {Burovski}, {Peterson}, {Weckesser},
  {Bright}, {van der Walt}, {Brett}, {Wilson}, {Jarrod Millman}, {Mayorov},
  {Nelson}, {Jones}, {Kern}, {Larson}, {Carey}, {Polat}, {Feng}, {Moore}, {Vand
  erPlas}, {Laxalde}, {Perktold}, {Cimrman}, {Henriksen}, {Quintero}, {Harris},
  {Archibald}, {Ribeiro}, {Pedregosa}, {van Mulbregt}, \&
  {Contributors}}]{Virtanen_2020}
{Virtanen}, P., {Gommers}, R., {Oliphant}, T.~E., {et~al.} 2020, Nature
  Methods, 17, 261

\bibitem[{{Werner} {et~al.}(2003){Werner}, {Deetjen}, {Dreizler}, {Nagel},
  {Rauch}, \& {Schuh}}]{Werner2003}
{Werner}, K., {Deetjen}, J.~L., {Dreizler}, S., {et~al.} 2003, in Astronomical
  Society of the Pacific Conference Series, Vol. 288, Stellar Atmosphere
  Modeling, ed. I.~{Hubeny}, D.~{Mihalas}, \& K.~{Werner}, 31

\bibitem[{{Werner} \& {Dreizler}(1999)}]{Werner1999}
{Werner}, K., \& {Dreizler}, S. 1999, Journal of Computational and Applied
  Mathematics, 109, 65

\bibitem[{{Worthey}(1994)}]{Worthey1994}
{Worthey}, G. 1994, \apjs, 95, 107

\bibitem[{{Zinn} {et~al.}(1972){Zinn}, {Newell}, \& {Gibson}}]{Zinn1972}
{Zinn}, R.~J., {Newell}, E.~B., \& {Gibson}, J.~B. 1972, \aap, 18, 390

\end{thebibliography}
\bibliographystyle{aasjournal}
\newpage

\appendix

\begin{table*}[!hbt]
\caption{pHB evolutionary model parameters from \citet{Moehler2019}}
\makebox[0.89\linewidth]
{
\begin{tabular}{cccccc}
\toprule
${\it M_{ZAHB}}$    & $T_{eff,ZAHB}$ & log $g_{ZAHB}$ & HB location & pHB behaviour                        & $M_{final}$   \\ 
(${\it M_{\odot}}$) & (K)            & ($cm s^{-2}$)  &             &                                          & (${\it M_{\odot}}$) \\
\hline
\multicolumn{6}{c}{{[}M/H{]} = $−$1, Age = 12 Gyr, ${\it M_{ZAMS}}$ = $0.85$ ${\it M_{\odot}}$}                                          \\ \hline
0.490          & 30138         & 5.74           & EHB         & post-EHB, no thermal pulses              & 0.490          \\
0.494         & 26706         & 5.52           & EHB         & 1 thermal pulse (like a LTP)             & 0.493         \\
0.495         & 26152         & 5.49           & EHB         & 1 thermal pulse (like a LTP)             & 0.495         \\
0.500           & 24040         & 5.34           & EHB         & 2 thermal pulses (like a LTP)            & 0.496         \\
0.530          & 17770         & 4.78           & BHB         & post-EAGB, 2 thermal pulses (like a LTP) & 0.499         \\
0.550          & 15264         & 4.48           & BHB         & TP-AGB + LTP                             & 0.504         \\
0.580          & 11097         & 3.80            & BHB         & TP-AGB + LTP                             & 0.513         \\
0.600           & 8815           & 3.35           & BHB         & TP-AGB                                   & 0.518         \\
0.650          & 5724           & 2.56           & RHB         & TP-AGB +LTP                              & 0.528*        \\
0.700           & 5484           & 2.50            & RHB         & TP-AGB                                   & 0.537         \\
0.750          & 5392           & 2.47           & RHB         & TP-AGB                                   & 0.545         \\
0.850          & 5315           & 2.46           & RHB         & TP-AGB                                   & 0.555         \\ \toprule
\multicolumn{6}{p{\textwidth}}{\textbf{Notes.} TP-AGB stands for thermally pulsing AGB, (V)LTP for (very) late thermal pulse.$^{(*)}$These sequences end up highly H-deficient due to burning or dilution of the H-rich envelope during the last He-shell flash.}
\end{tabular}
}
\label{tab:evol model params}
\end{table*}


\begin{table*}[!hbt]
\caption{The UVIT photometry of pHB member stars in the GC, NGC 2808. Column 1 lists the star ID; columns 2 and 3 correspond to the R.A. and Decl. of the stars; columns 4 to 15 give the magnitudes and errors (AB system) in different UVIT filters. Note that the magnitudes are not corrected for extinction. The stars with IDs 12 to 27 are not resolved by UVIT and hence not available in this table. }
\makebox[0.79\linewidth]
{
\begin{tabular}{ccccccccccccccc}
\toprule
ID & R.A. & Decl. & F154W & err1 & F169M & err2 & N242W & err3 & N245M & err4 & N263M & err5 & N279N & err6 \\\
 & (deg) & (deg) & & & & & & & & & & & & \\
\toprule
Star 1  & 137.96181 & -64.84165 & 18.437 & 0.039 & 18.514 & 0.041 & 19.017 & 0.033 & 18.924 & 0.070 & 18.757 & 0.093 & 18.783 & 0.083 \\
Star 2  & 138.02575 & -64.84307 & 16.349 & 0.018 & 16.346 & 0.023 & 16.923 & 0.029 & 16.549 & 0.023 & 16.389 & 0.045 & 16.515 & 0.031 \\
Star 3  & 138.06046 & -64.86146 & 16.376 & 0.020 & 16.417 & 0.018 & 17.450 & 0.038 & 17.135 & 0.027 & 17.077 & 0.056 & 17.190 & 0.037 \\
Star 4  & 138.05040 & -64.84379 & 15.899 & 0.018 & 15.891 & 0.018 & 16.724 & 0.031 & 16.380 & 0.028 & 16.325 & 0.040 & 16.444 & 0.032 \\
Star 5  & 138.04719 & -64.87492 & 18.313 & 0.030 & 18.289 & 0.034 & 19.015 & 0.039 & 18.710 & 0.052 & 18.476 & 0.088 & 18.483 & 0.068 \\
Star 6  & 138.06813 & -64.86459 & 17.817 & 0.025 & 17.727 & 0.031 & 17.826 & 0.024 & 17.680 & 0.039 & 17.419 & 0.053 & 17.439 & 0.055 \\
Star 7  & 138.05264 & -64.86779 & 17.888 & 0.028 & 17.825 & 0.029 & 18.629 & 0.045 & 18.274 & 0.052 & 18.105 & 0.094 & 18.224 & 0.060 \\
Star 8  & 138.04842 & -64.84887 & 17.597 & 0.025 & 17.559 & 0.029 & 18.087 & 0.031 & 17.702 & 0.046 & 17.473 & 0.064 & 17.525 & 0.049 \\
Star 9  & 138.04760 & -64.85790 & 17.657 & 0.025 & 17.611 & 0.034 & 18.208 & 0.055 & 17.799 & 0.041 & 17.588 & 0.051 & 17.524 & 0.056 \\
Star 10 & 138.03820 & -64.87028 & 17.613 & 0.023 & 17.579 & 0.030 & 18.743 & 0.061 & 18.031 & 0.042 & 17.930 & 0.075 & 17.856 & 0.059 \\
Star 11 & 137.96088 & -64.85092 & 18.006 & 0.032 & 18.004 & 0.035 & 18.540 & 0.032 & 18.277 & 0.044 & 17.995 & 0.071 & 18.403 & 0.065 \\
Star 28 & 137.85352 & -64.87929 & 18.184 & 0.029 & 18.178 & 0.044 & 18.652 & 0.030 & 18.525 & 0.048 & 18.342 & 0.087 & 18.623 & 0.087 \\
Star 29 & 137.92453 & -64.70158 & 17.404 & 0.025 & 17.465 & 0.025 & 17.807 & 0.022 & 17.749 & 0.034 & 17.675 & 0.048 & 17.814 & 0.054 \\
Star 30 & 138.08369 & -64.84865 & 17.747 & 0.025 & 17.738 & 0.031 & 18.173 & 0.026 & 18.010 & 0.041 & 17.975 & 0.071 & 17.932 & 0.053 \\
Star 31 & 138.00728 & -64.79293 & 16.723 & 0.020 & 16.748 & 0.020 & 17.058 & 0.019 & 16.917 & 0.020 & 16.855 & 0.042 & 16.950 & 0.039 \\
Star 32 & 137.99974 & -64.89071 & 16.906 & 0.025 & 16.953 & 0.024 & 17.844 & 0.035 & 17.583 & 0.038 & 17.515 & 0.044 & 17.712 & 0.066 \\
Star 33 & 138.02306 & -64.89646 & 18.447 & 0.034 & 18.477 & 0.036 & 18.759 & 0.025 & 18.569 & 0.055 & 18.339 & 0.082 & 18.651 & 0.093 \\
Star 34 & 138.09187 & -64.87727 & 17.489 & 0.025 & 17.493 & 0.027 & 18.003 & 0.035 & 17.779 & 0.042 & 17.745 & 0.064 & 17.704 & 0.063 \\
\toprule
\end{tabular}
}
\label{pHB UVIT mag}
\end{table*}


\begin{figure}[!hbt]
\centering
\subfloat{\includegraphics[width=0.5\columnwidth, height = 8cm]{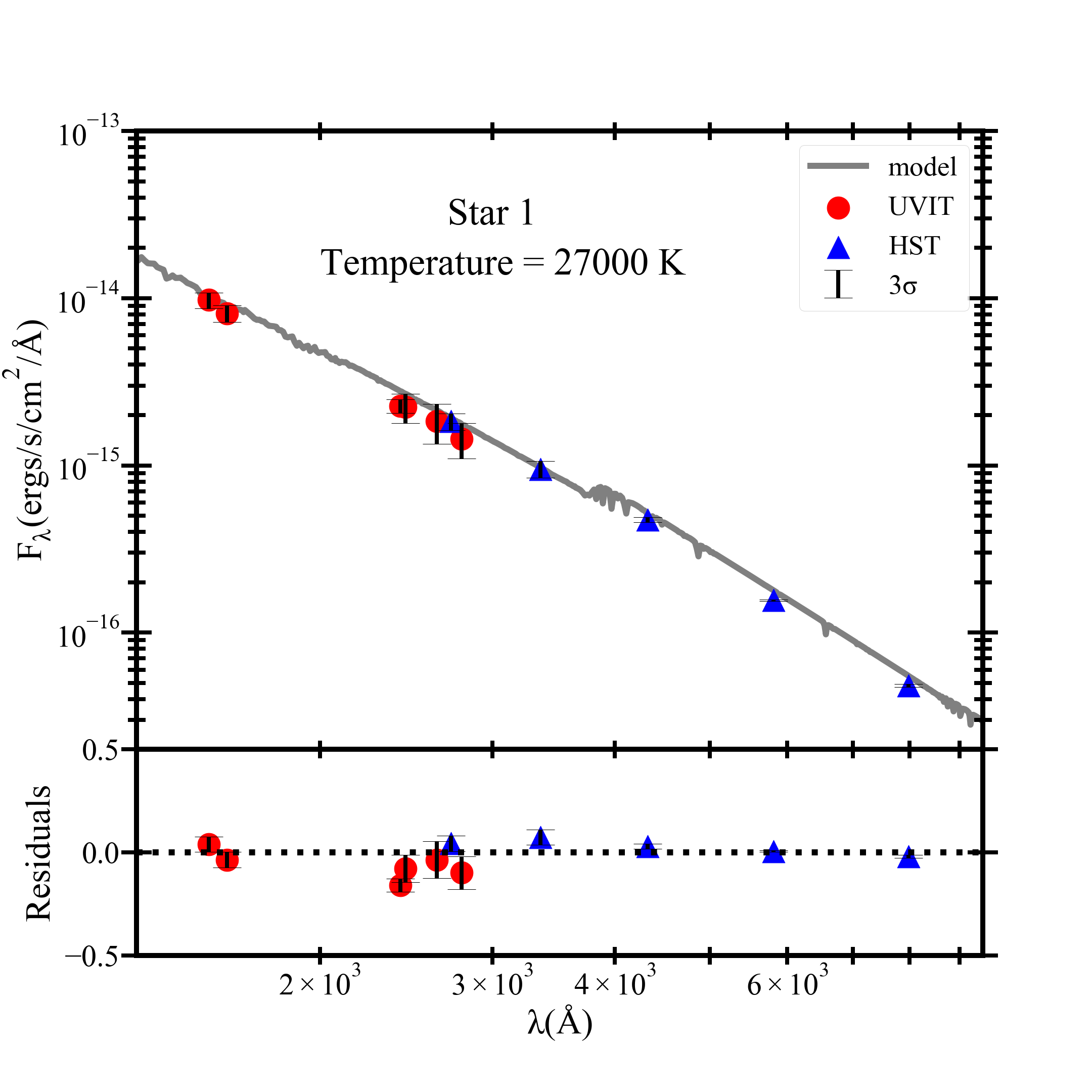}}
\subfloat{\includegraphics[width=0.5\columnwidth, height = 8cm]{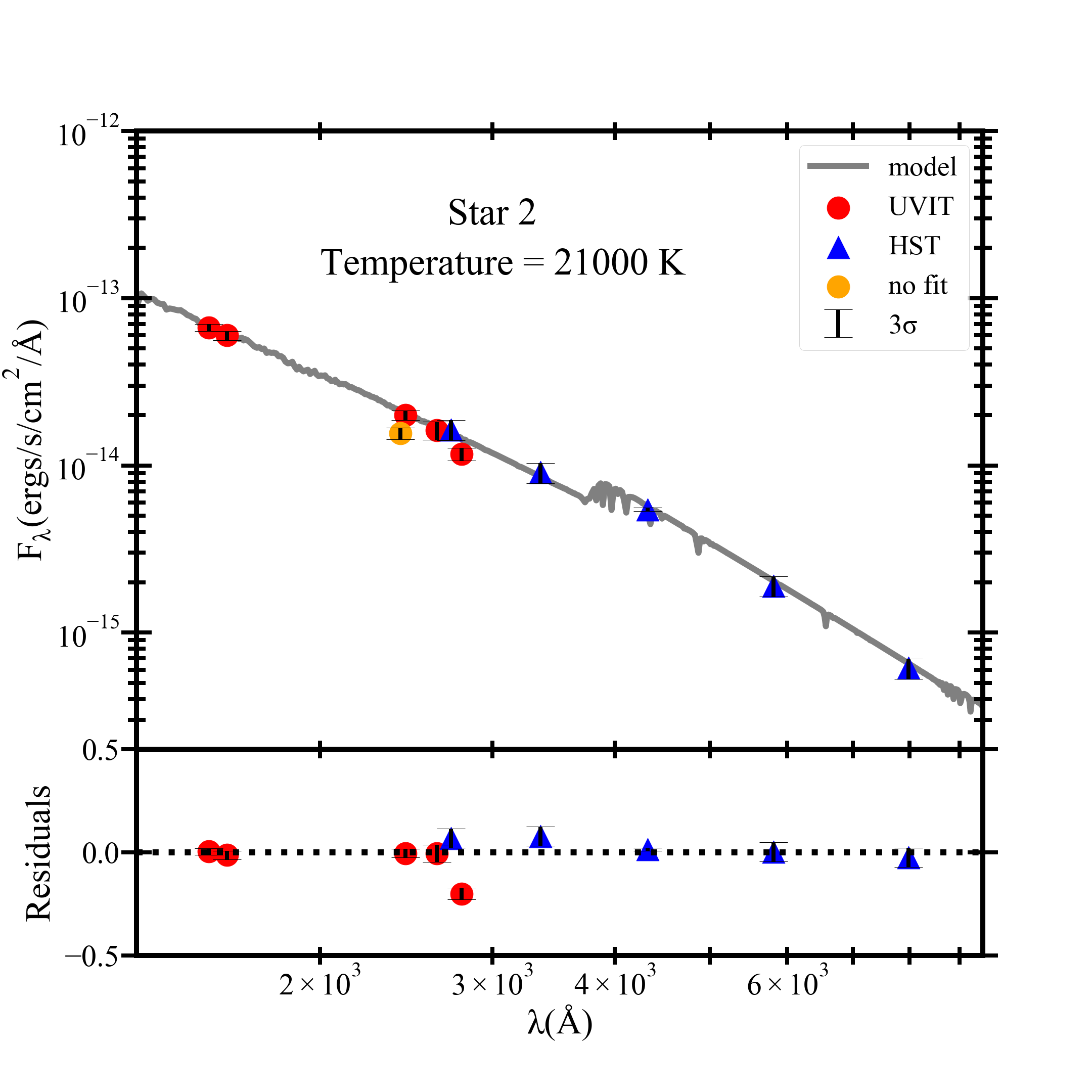}}\\[-2ex]
\subfloat{\includegraphics[width=0.5\columnwidth, height = 8cm]{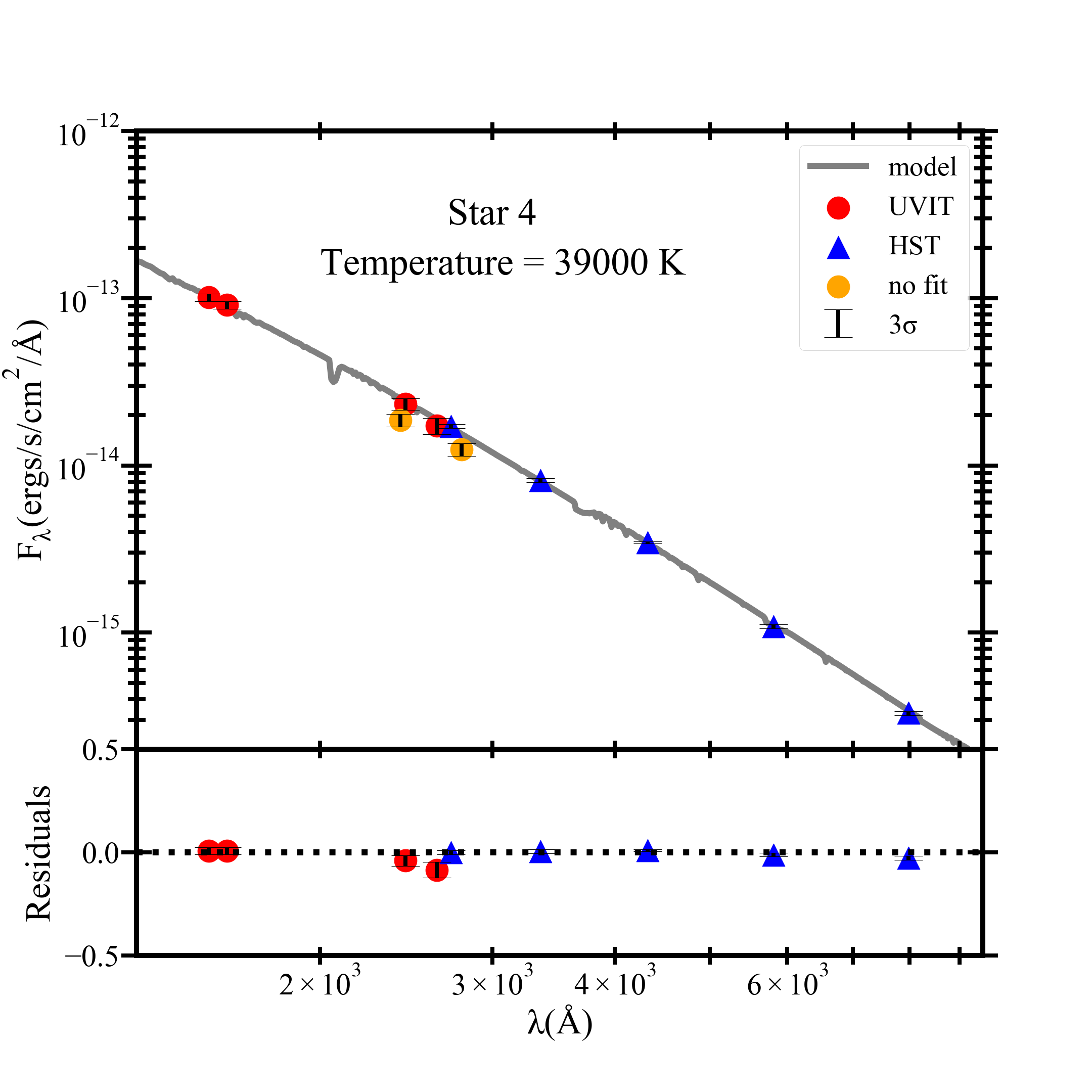}}
\subfloat{\includegraphics[width=0.5\columnwidth, height = 8cm]{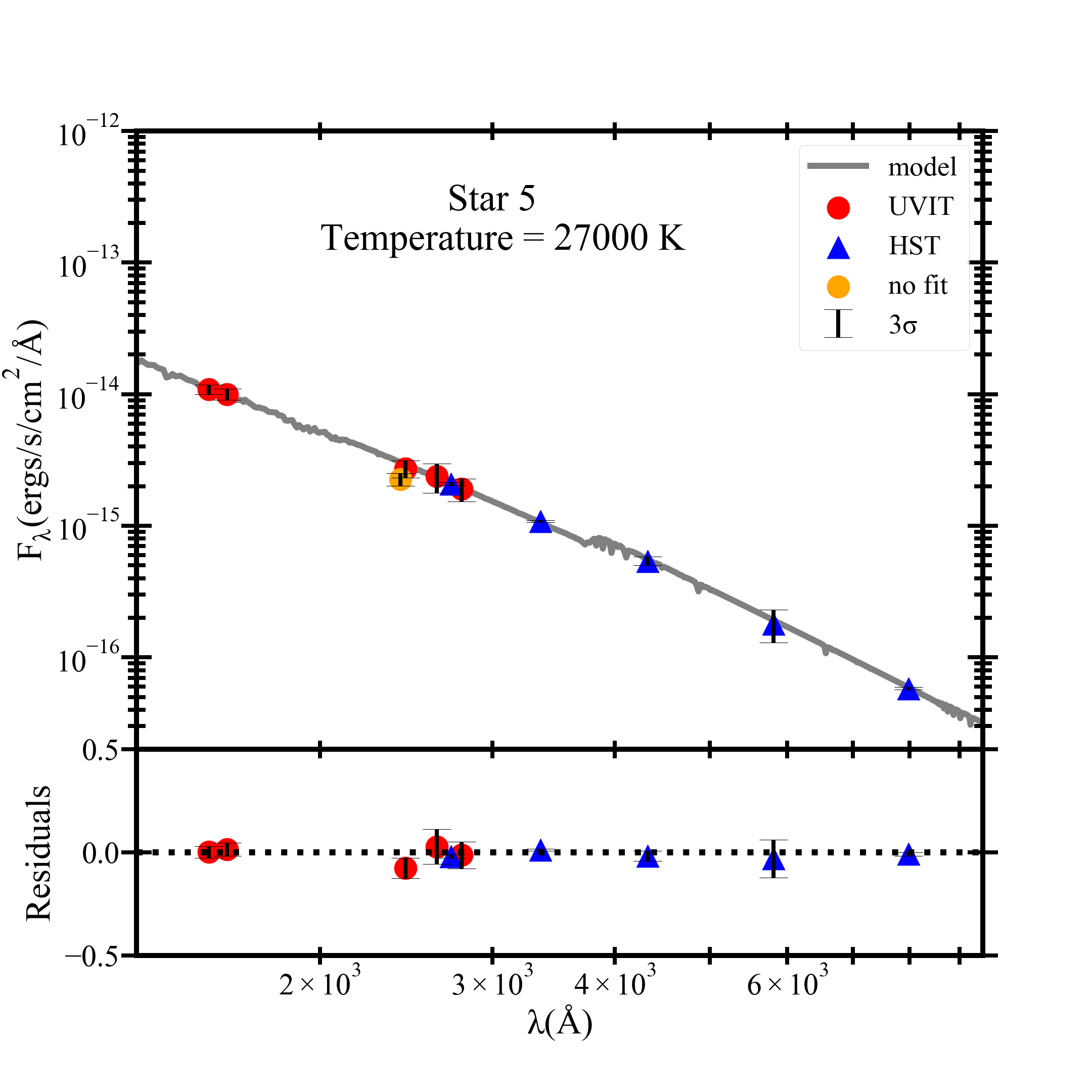}}
\caption{SEDs for UV-bright stars listed in Table \ref{pHB UVIT mag} except those of Star 3 and Star 28 which are already shown in Figure \ref{SED_eg}. For stars lying within the {\it HST} FOV, photometric data from UVIT and {\it HST} are used for fitting the SEDs. For those in the outer region, data from UVIT, {\it GALEX}, {\it Gaia} and ground-based optical data from \citet{Stetson2019} are used.  The photometric points excluded from the fitting procedure are shown with orange symbols. The gray line shows the model spectrum. The residuals of SED fit are shown in the bottom panels of every plot.}
\label{seds_uvit_phb}
\end{figure}


\begin{figure}
\ContinuedFloat
\centering
\subfloat{\includegraphics[width=0.5\columnwidth, height = 8cm]{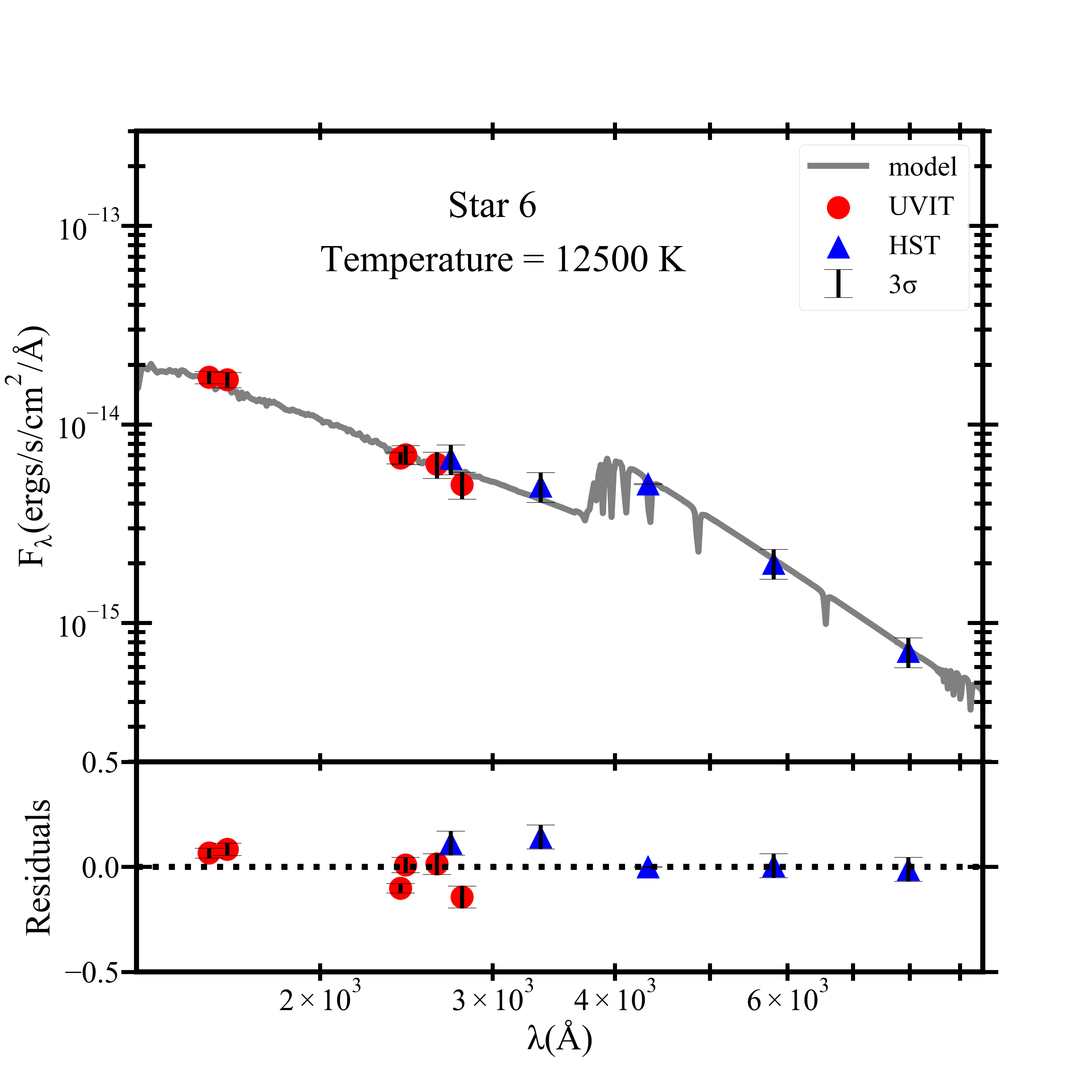}}
\subfloat{\includegraphics[width=0.5\columnwidth, height = 8cm]{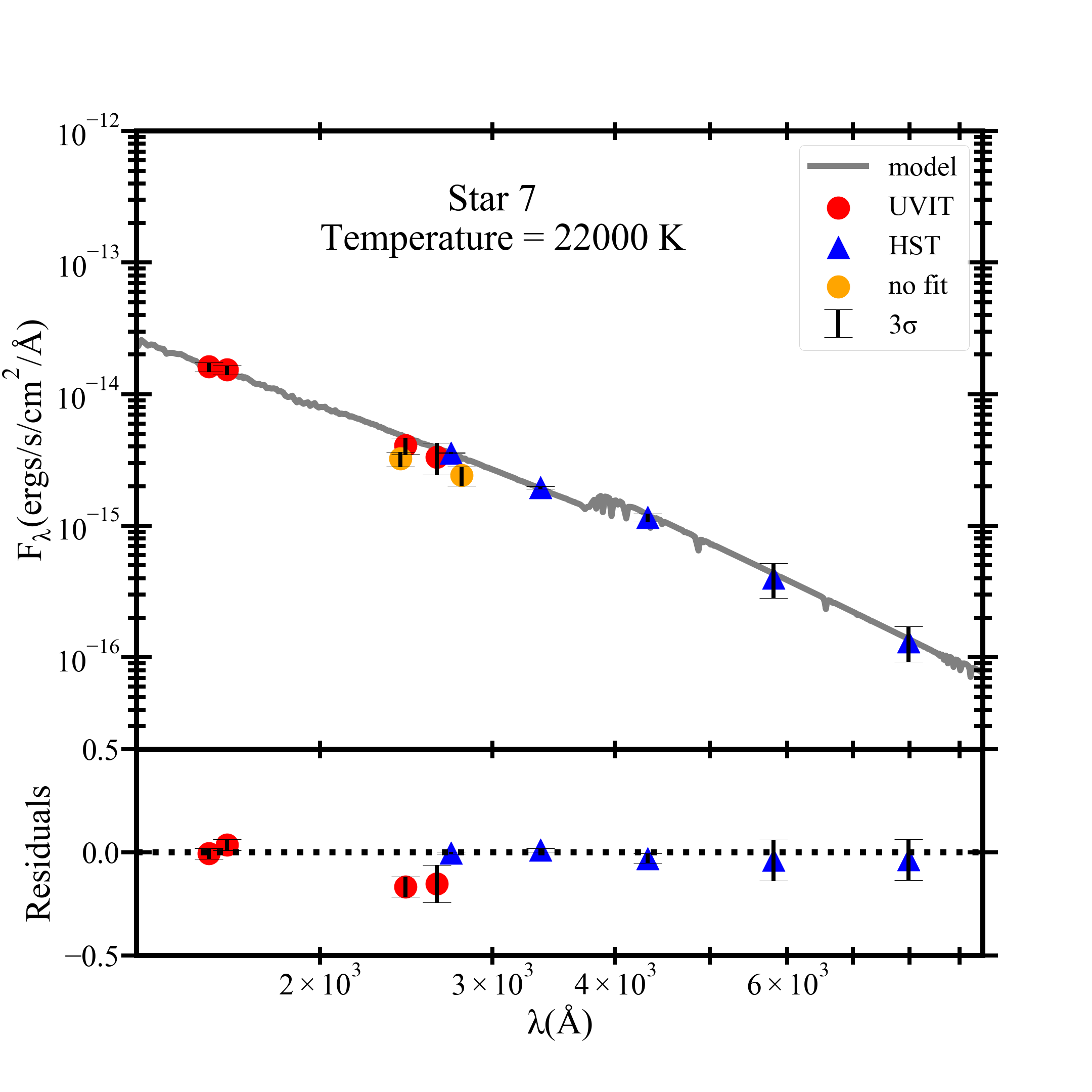}}\\[-2ex]
\subfloat{\includegraphics[width=0.5\columnwidth, height = 8cm]{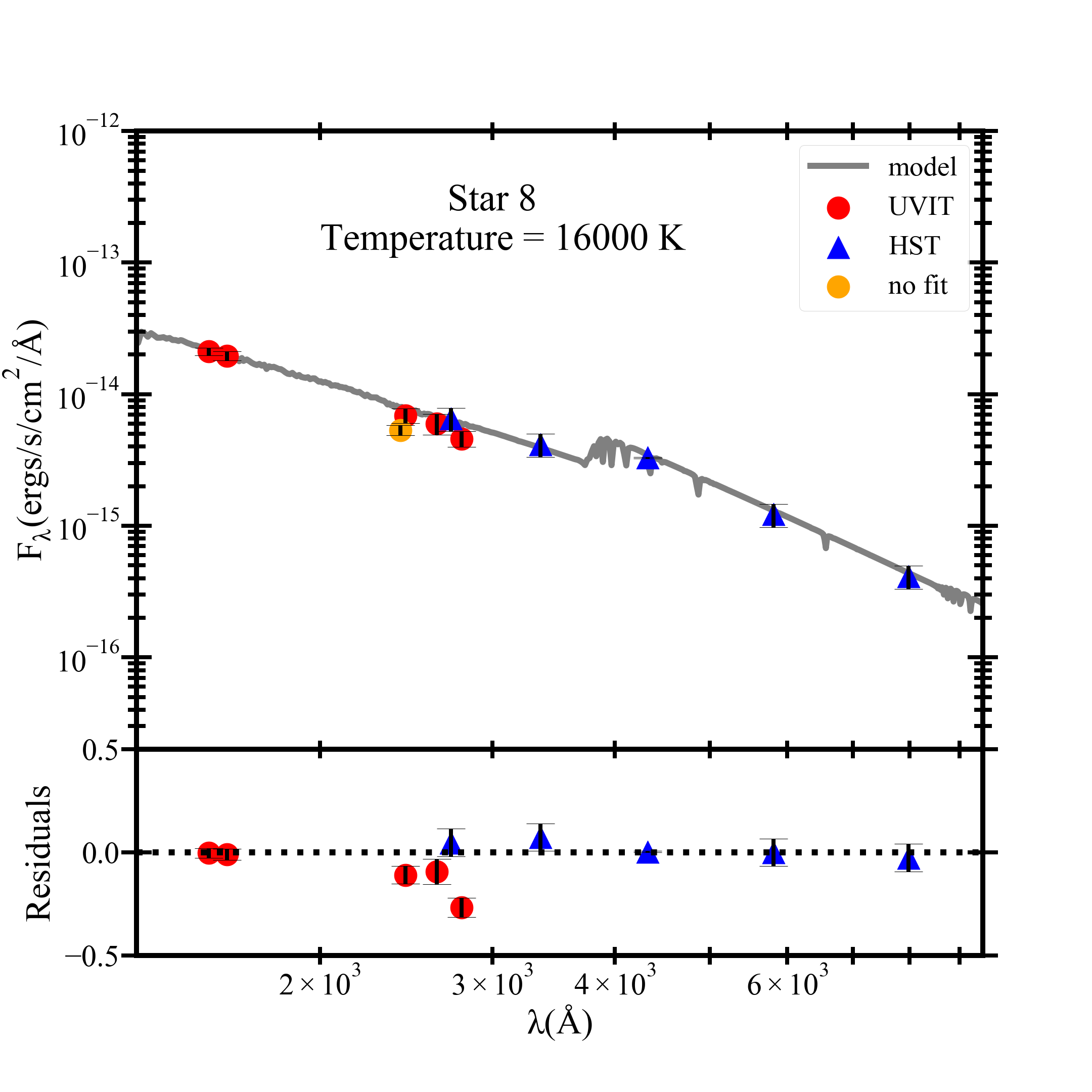}}
\subfloat{\includegraphics[width=0.5\columnwidth, height = 8cm]{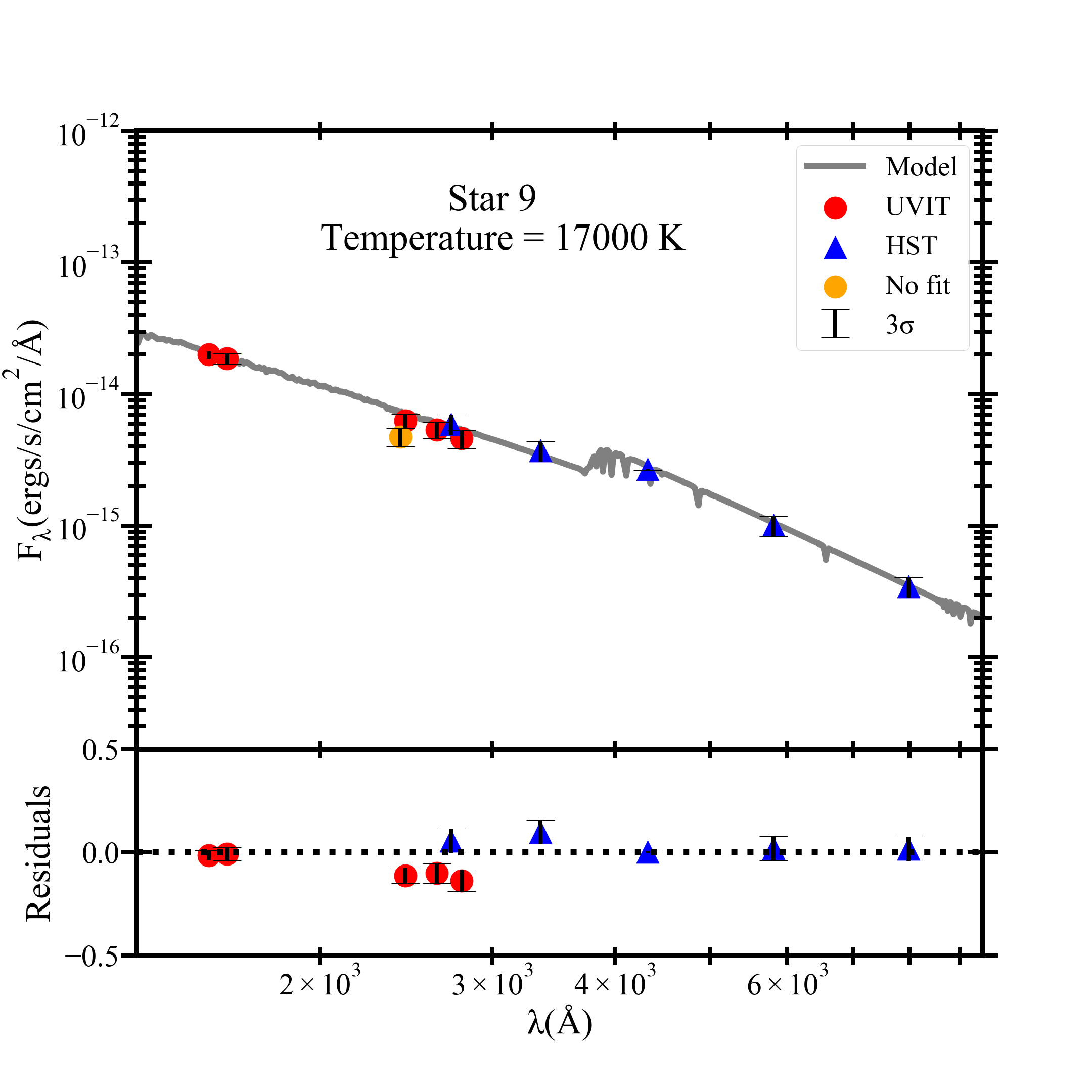}}
\caption{(continued.)}
\end{figure}


\begin{figure}
\ContinuedFloat
\centering
\subfloat{\includegraphics[width=0.5\columnwidth, height = 8cm]{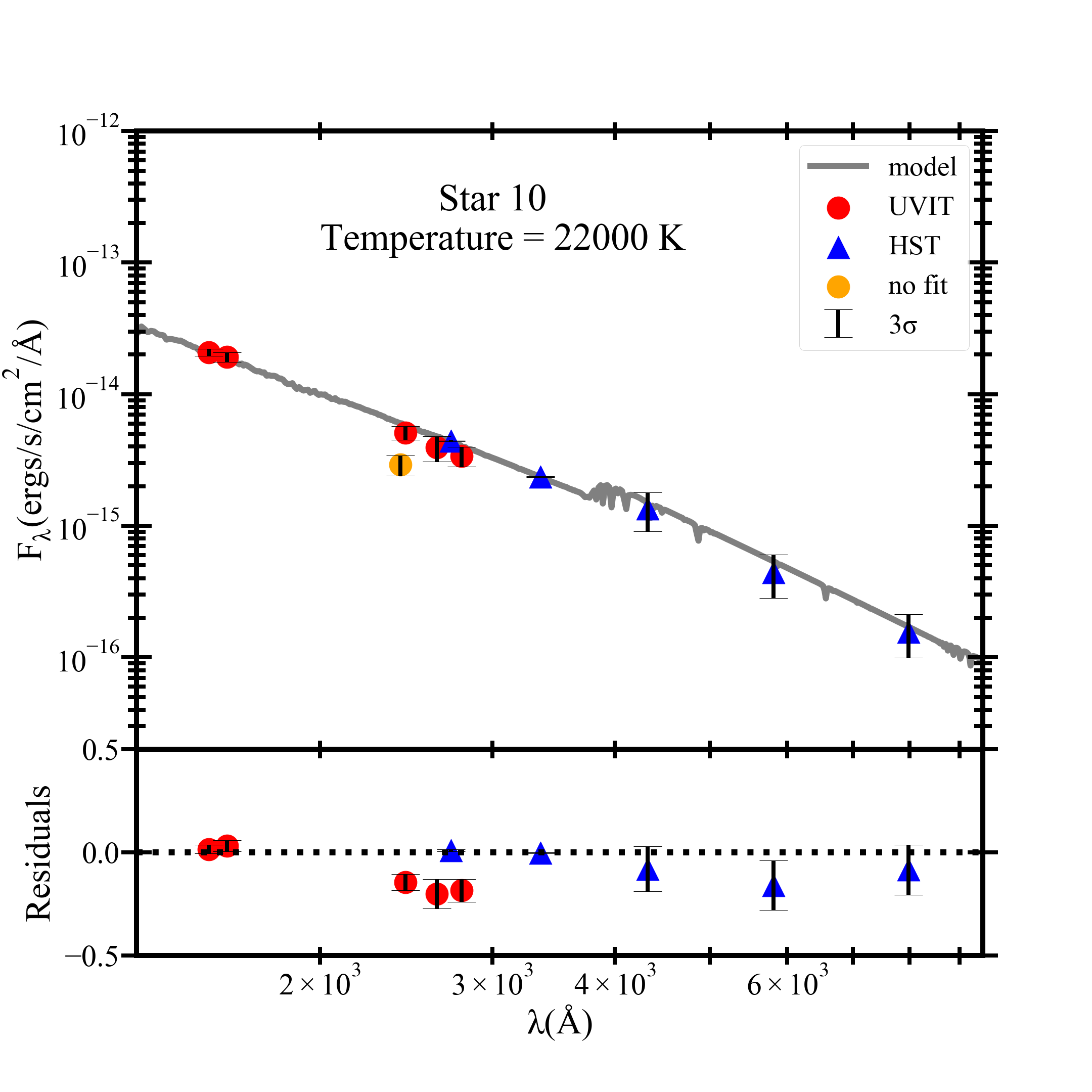}}
\subfloat{\includegraphics[width=0.5\columnwidth, height = 8cm]{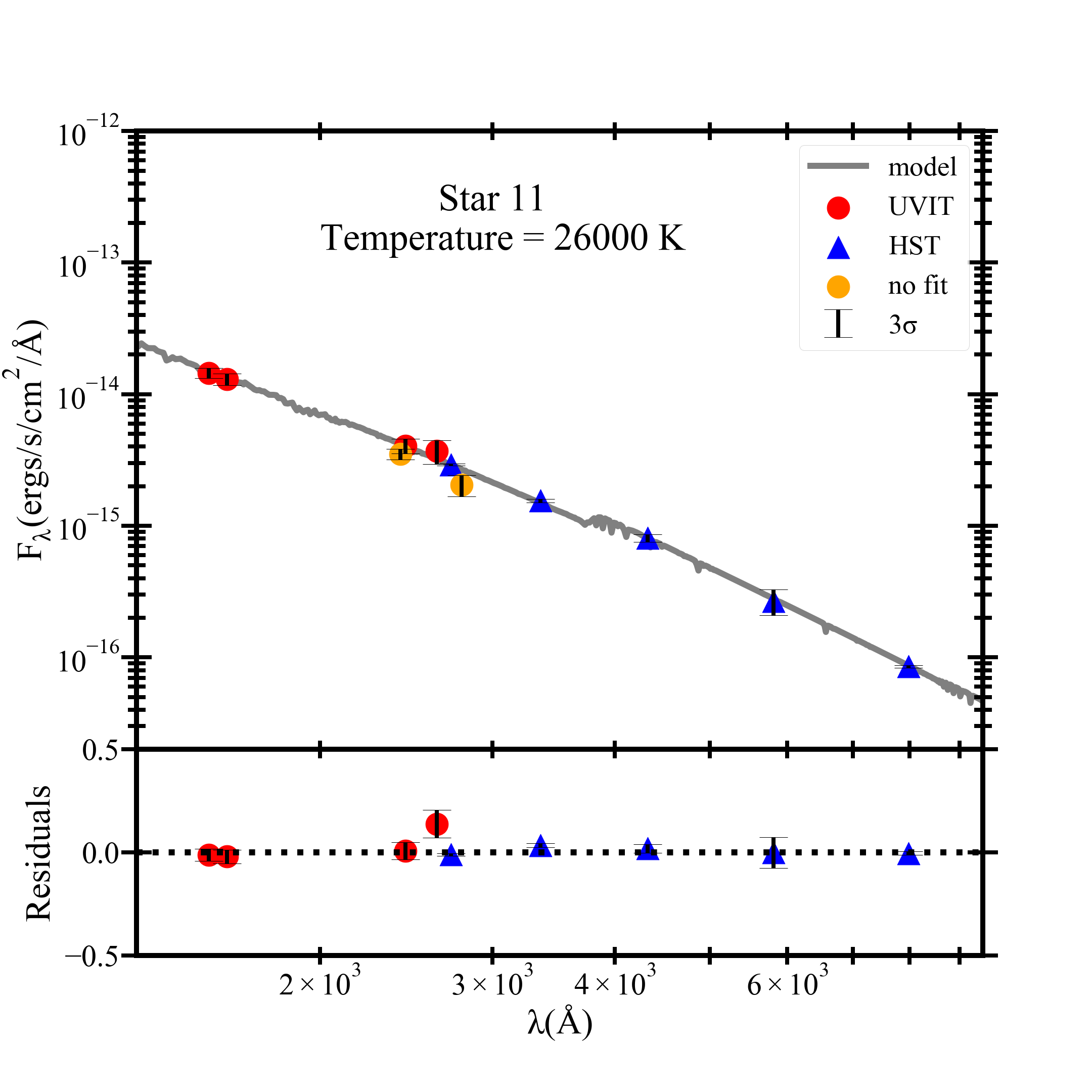}}\\[-2ex]
\subfloat{\includegraphics[width=0.5\columnwidth, height = 8cm]{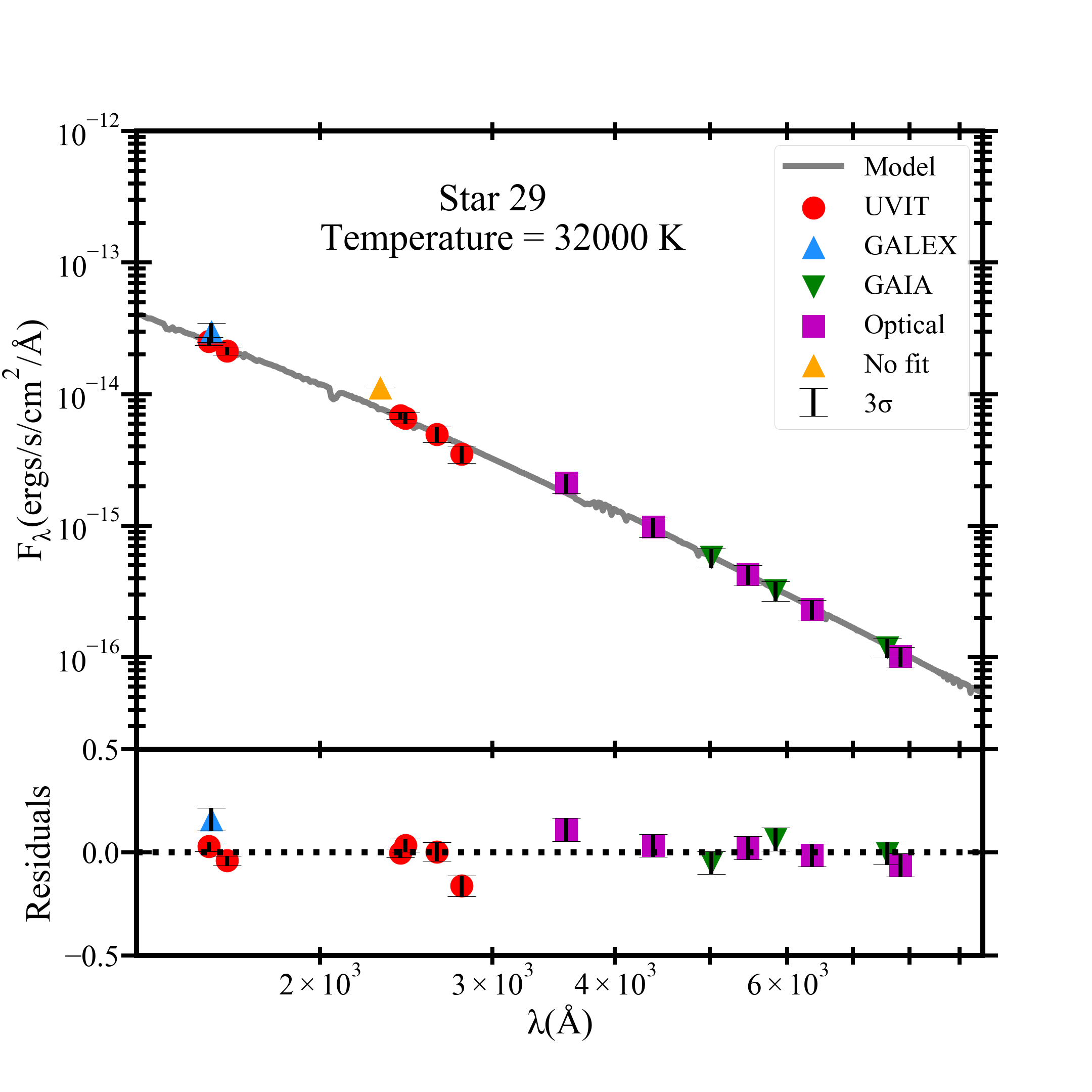}}
\subfloat{\includegraphics[width=0.5\columnwidth, height = 8cm]{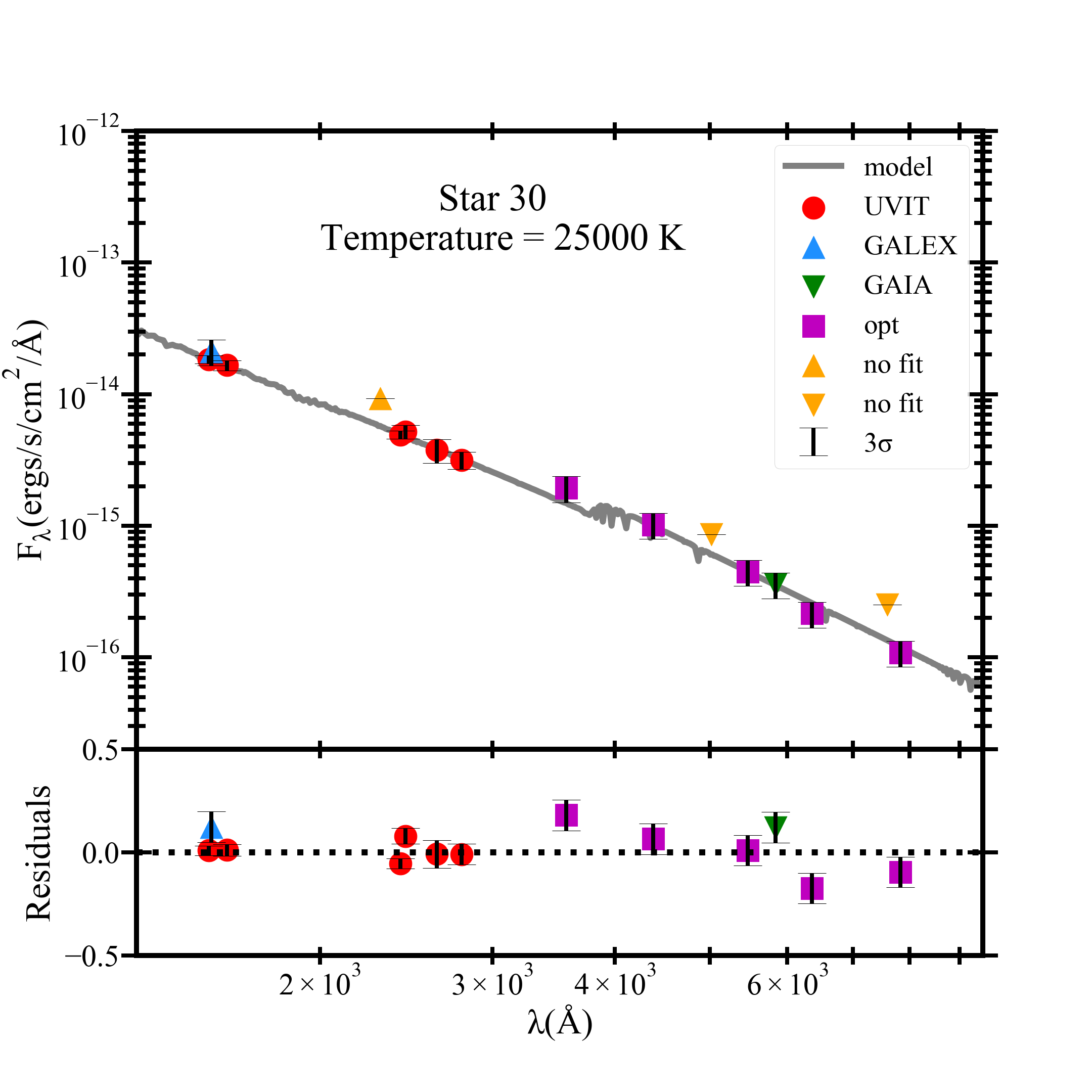}}
\caption{(continued.)}
\end{figure}


\begin{figure}
\ContinuedFloat
\centering
\subfloat{\includegraphics[width=0.5\columnwidth, height = 8cm]{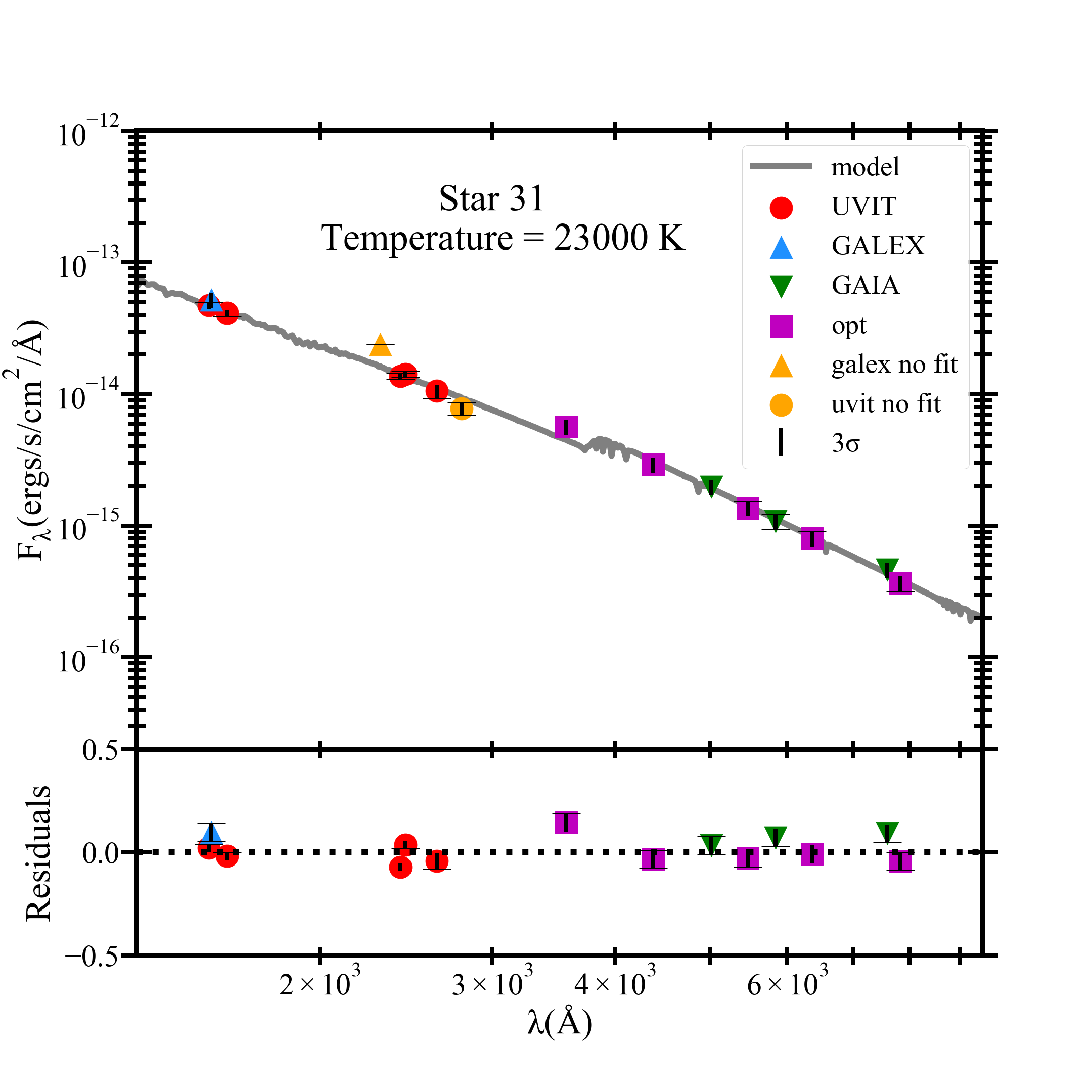}}
\subfloat{\includegraphics[width=0.5\columnwidth, height = 8cm]{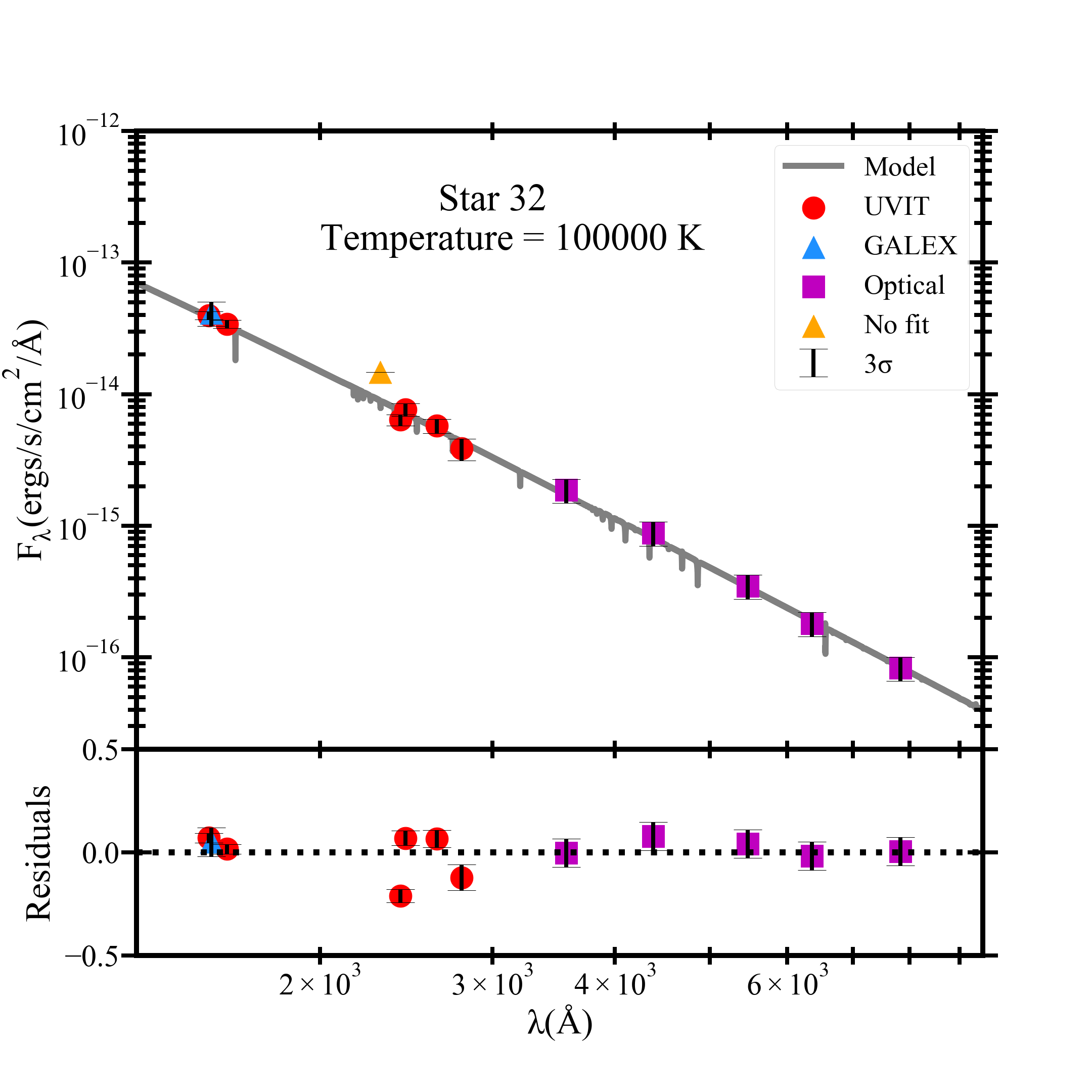}}\\[-2ex]
\subfloat{\includegraphics[width=0.5\columnwidth, height = 8cm]{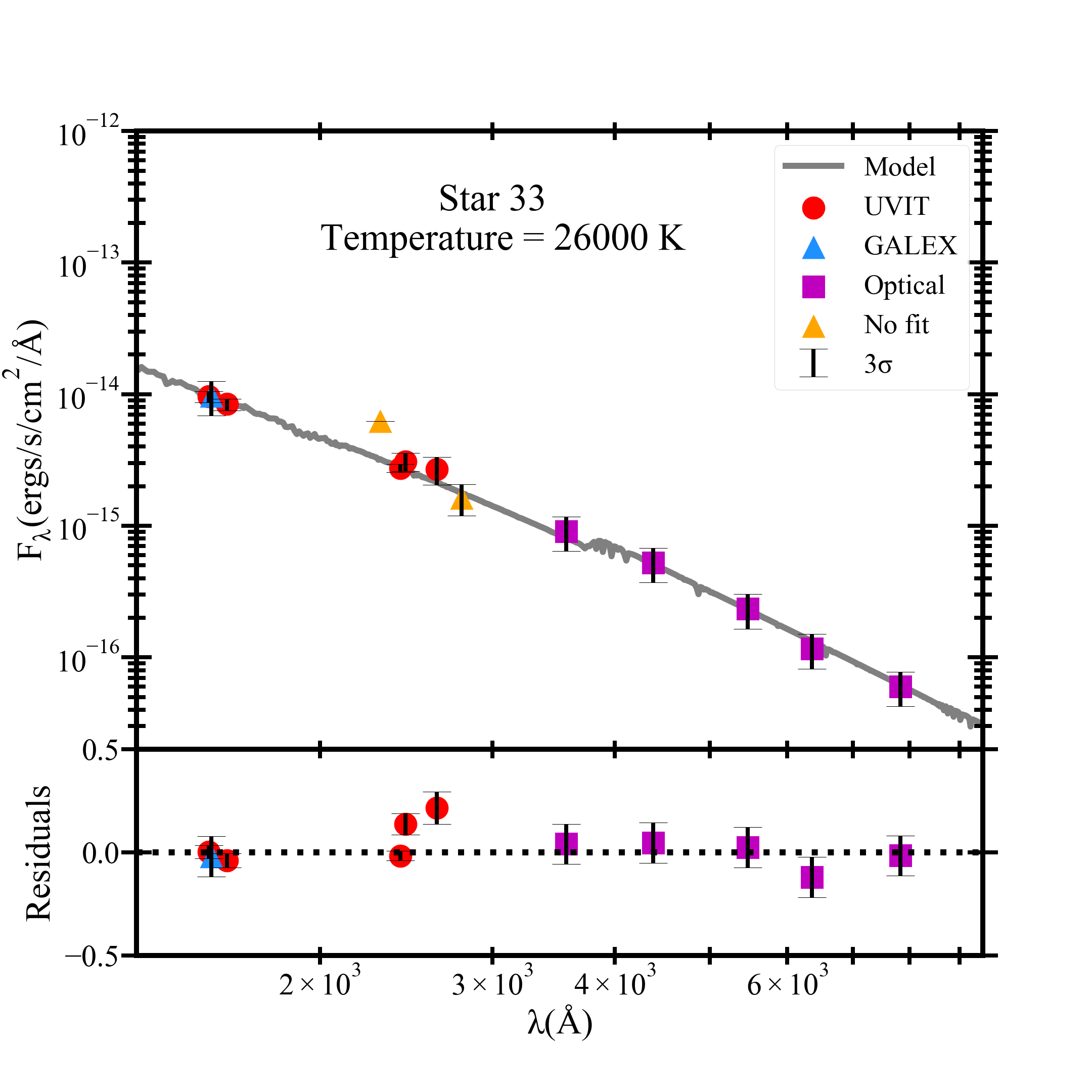}}
\subfloat{\includegraphics[width=0.5\columnwidth, height = 8cm]{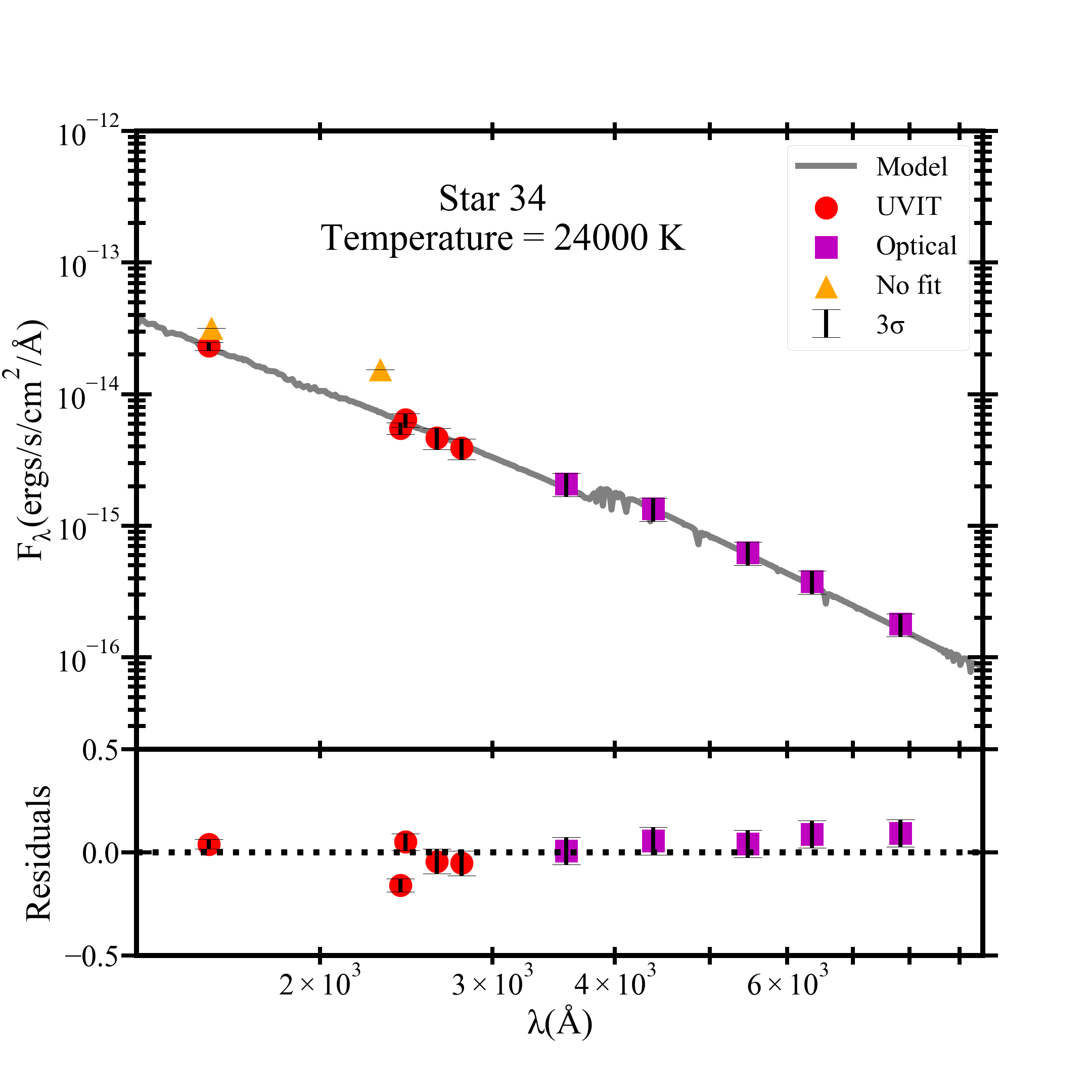}}
\caption{(continued.)}
\end{figure}

\end{document}